\documentclass[11pt]{article}
\usepackage{amsmath}
\usepackage{amsthm}
\usepackage{cancel}
\usepackage{amssymb}
\usepackage{listings}  
\usepackage{algorithm}
\usepackage{subfig}
\usepackage{color}
\usepackage[english]{babel}
\usepackage{graphicx}
\usepackage{wrapfig,epsfig}
\usepackage{epstopdf}
\usepackage{multirow}
\usepackage{url}
\usepackage{graphicx}
\usepackage{color}
\usepackage{epstopdf}
\usepackage{algpseudocode}
\usepackage{scrextend}
\usepackage[T1]{fontenc}
\usepackage{bbm}
\usepackage{comment}
\usepackage{animate}
\usepackage{bbold}
\usepackage{multicol}{}
\usepackage{wasysym}
\DeclareMathAlphabet{\mathpzc}{OT1}{pzc}{m}{it}

\usepackage{tikz}
\usepackage{hyperref}
\hypersetup{colorlinks=true,citecolor=red,linkcolor=blue}
\usetikzlibrary{arrows}
\usepackage[margin=1in]{geometry}
\graphicspath{{./figs/}}

\usepackage{mathtools}

\newtheorem{theorem}{Theorem}[section]
\newtheorem{lemma}[theorem]{Lemma}
\newtheorem{definition}[theorem]{Definition}

\newtheorem{proposition}[theorem]{Proposition}

\newtheorem{observation}[theorem]{Observation}

\newtheorem{remark}[theorem]{Remark}
\newtheorem{claim}[theorem]{Claim}

\newtheorem{question}[theorem]{Question}

\newcommand{\wh}{\widehat}
\newcommand{\wt}{\widetilde}

\newcommand{\eps}{\epsilon}

\newcommand{\R}{\mathbb{R}}

\renewcommand{\varepsilon}{\epsilon}
\renewcommand{\tilde}{\wt}
\renewcommand{\hat}{\wh}
\renewcommand{\eps}{\epsilon}
\renewcommand{\d}{\mathrm{d}}

\newcommand{\cH}{{\cal H}}
\newcommand{\one}[1]{{\mathbb 1}\left[{#1}\right]}

\DeclareMathOperator*{\E}{{\mathbb{E}}}

\DeclareMathOperator{\poly}{poly}

\DeclareMathOperator{\dia}{dia}

\DeclareMathOperator{\dis}{dis}

\DeclareMathOperator{\avg}{avg}

\DeclareMathOperator{\argmin}{argmin}
\DeclareMathOperator{\Vol}{Vol}

\makeatletter
\newcommand*{\RN}[1]{\expandafter\@slowromancap\romannumeral #1@}
\makeatother

\usepackage{lineno}

\title{Sub-quadratic $(1+\eps)$-approximate Euclidean Spanners, with Applications}

\author{
Alexandr Andoni\thanks{\texttt{andoni@cs.columbia.edu}. Columbia University.}
\quad
Hengjie Zhang\thanks{\texttt{hengjie.z@columbia.edu}. Columbia University.}
}

\begin{document}

\maketitle

\begin{abstract}
We study graph spanners for point-set in the high-dimensional Euclidean space. On the one hand, we prove that spanners with stretch $<\sqrt{2}$ and subquadratic size are not possible, even if we add Steiner points. On the other hand, if we add extra nodes to the graph (non-metric Steiner points), then we can obtain $(1+\eps)$-approximate spanners of subquadratic size. We show how to construct a spanner of size $n^{2-\Omega(\eps^3)}$, as well as a directed version of the spanner of size $n^{2-\Omega(\eps^2)}$.

We use our directed spanner to obtain an algorithm for computing $(1+\eps)$-approximation to Earth-Mover Distance (optimal transport) between two sets of size $n$ in time $n^{2-\Omega(\eps^2)}$.
\end{abstract}

\section{Introduction}
A spanner for a metric defined by the graph $G$ is a subgraph $H$ such that all shortest path distances in $H$ approximate shortest path distances in $G$ up to some approximation $\alpha$. Originally introduced in \cite{peleg1989graph}, the notion has been very influential in theory and practice --- see, for example, the recent survey \cite{ahmed2020graph} and earlier surveys \cite{eppstein1999spanning, zwick2001exact}. The classic result is that, for any integer $k\ge1$, for any weighted graph $G$, one can construct a spanner of size $O(n^{1+1/k})$ approximating all distances up to factor $2k-1$ \cite{peleg1989graph}, termed stretch. This is tight under the Erd\"os girth conjecture~\cite{Erd63}.

Graph spanners have been studied in a number of settings and variants. While we refer the reader to the survey \cite{ahmed2020graph} for the vast list of relevant papers, we highlight that there are a number of parameters/versions of spanners considered: approximation (e.g., multiplicative, additive, and others), which pairs to preserve, properties of the graph $G$, computation time, data structures (distance oracles), etc.

An almost disjoint line of research concerns {\em geometric spanners}, where the graph $G$ is defined implicitly as the pairwise distance in some metric $M$; introduced in \cite{chew1986there} even before spanners where defined. In particular, here $G$ is a complete graph, where each pair of nodes is connected with an edge with weight equal to their distance. The classic results here show that an $n$-point metric in the Euclidean space $\R^d$ admits a $(1+\eps)$-approximation spanner of size $O(n\cdot \eps^{-d+1})$ \cite{clarkson1987approximation, keil1988approximating, keil1992classes, althofer1993sparse}; see also the survey \cite{narasimhan2007geometric}. 

More recent progress \cite{le2022truly,bhore2022euclidean} showed tightness of the bound $n\eps^{-d+1}$, but also that {\em Steiner} spanners can improve the size to $O(n\eps^{\tfrac{-d+1}{2}})$. A Steiner spanner is a graph whose nodes are the pointset $P\subset \R^d$ together with some extra Steiner points $\in \R^d$. Whenever an edge is added to the spanner, its weight is the direct distance between the corresponding points (so that no pair is shortcut).

In contrast to the two extremely well-studied settings from above, the {\em high-dimensional} geometric spanners have surprisingly received much less attention. In particular, we are aware of only one (theoretical) result: \cite{his13} who obtained $c$-spanner of size $O(n^{1+O(1/c^2)})$ for any $c\ge 3.4$. This is despite the ubiquity of high-dimensional datasets and
the fact that many (massive) graphs are actually ``similarity graphs''
\footnote{While similarity graph is a bit different from the aforementioned complete graph --- e.g., the edges are only for pairs of points at distance $\le r$ --- the tools from here apply to such graphs nonetheless.}, obtained by connecting close points of some dataset $P\in\R^d$ --- subsequently used for various clustering applications; see recent \cite{dhulipala2022hierarchical} and references therein. Indeed, the spanner from \cite{his13} has been implemented in \cite{carey2022stars}, yielding orders of magnitude speed up in graph building, without affecting down-the-stream clustering which leverages the spanner only.

To put our lack of understanding into perspective, we don't even know if we can obtain subquadratic-size spanners for approximation $c\le 3.4$.

\subsection{Our contributions}

We build new high-dimensional geometric spanners in the $1+\eps$ approximation regime, asking the following main question:
\begin{question}
\label{q:main}
Do there exist spanners for high-dimensional Euclidean pointset with $1+\eps$ approximation and subquadratic size?
\end{question}

From now on, we assume we have an $n$-point dataset $P\in\R^d$ under the Euclidean distance, where $d\gg \log n$. For a graph $H$, we use $\d_H(p,q)$ to denote the shortest path distance between $p$ to $q$ in $H$. We also let $\Delta$ be the aspect ratio of $P$: $\Delta=\tfrac{\max_{p,q\in P} \|p-q\|_2}{\min_{p,q\in P, p\neq q} \|p-q\|_2}$.

\paragraph{Lower bound for spanners.} We first give indication that it is impossible to obtain $(1+\eps)$ approximation with a sub-quadratic-sized spanner even if we allow Steiner points. In particular, we prove the following two lower bounds for the {\em Hamming} metric and Euclidean metric, respectively.

\begin{theorem}[See Section~\ref{sec:lb_hamming}]\label{thm:lower_bound_intro}
    For any $\eps>0$, there exists a dataset $P\subset \{0,1\}^d$ such that the following holds. Fix a graph $H$ with vertices identified with $P$, such that for any $p,q\in P$, 
    $$\|p-q\|_1\le \d_H(p,q) < (2-\eps) \cdot \|p-q\|_1.
    $$
    Then $H$ must have $\Omega(n^2\eps^4)$ edges. This remains true even if $H$ contains Steiner vertices corresponding to points in $\{0,1\}^d$.
\end{theorem}

\begin{theorem}[See Section~\ref{sec:lb_euclidean}]\label{thm:lower_bound_l2_intro}
    For any $\eps>0$, there exists a dataset $P\subset S^{d-1}(0,1)$ such that the following holds. Fix a graph $H$ with vertices identified with $P$, such that for any $p,q\in P$, 
    $$\|p-q\|_2\le \d_H(p,q) < (\sqrt{2}-\eps) \cdot \|p-q\|_2.
    $$
    Then $H$ must have $\Omega(n^2\eps^4/\log^2n)$ edges. This remains true even if $H$ contains Steiner vertices corresponding to points in $\R^d$.
\end{theorem}

\paragraph{Spanners with non-metric Steiner nodes.} Next, we show that if we consider a more general class of Steiner spanners, where we allow graph nodes that do not correspond to points $\in \R^d$, we can indeed obtain $(1+\eps)$-spanners of subquadratic size, answering positively the question from above. We refer to such spanners as {\em non-metric} Steiner spanners.

Furthermore, we distinguish two types of non-metric Steiner spanners: undirected, and directed. In particular, a directed spanner $H$ for sets $P,Q$ with stretch $c>1$ is a directed graph with vertex-set $P\cup Q\cup S$, satisfying for all $p\in P,q\in Q$:
$$
\|p-q\|\le \vec{\d}_H(p,q)\le c\|p-q\|,
$$
where $\vec{\d}_H(p,q)$ as the shortest path from $p\in P$ to $q\in Q$. When we have one dataset $P$, we take $Q$ to be a copy of $P$. Note that directed spanner only makes sense for non-metric Steiner version.

\begin{theorem}[See Section~\ref{sec:directedSpanner}]\label{thm:upper_bound_intro}
    Fix $\eps\in(0,1)$. For any dataset $P\in \R^d$, there exists a directed graph $H$ on vertices $P\cup S$, such that for any $p,q\in P$, 
    $$
    \|p-q\|_2\le \vec{\d}_H(p,q) < (1+\eps) \cdot \|p-q\|_2.
    $$
    Furthermore, the number of edges of $H$, the size of $S$, and the construction time are all bounded by $O(n^{2-\Omega(\eps^2)}\log\Delta)$, where $\Delta$ is the aspect ratio of $P$.
\end{theorem}

While the directed spanner turns out to be already sufficient for our application below, there other applications which may require an undirected graph (as many graph problems are easier for undirected graphs). Thus, we also show that one can construct an {\em undirected} spanner, albeit the parameters are less efficient than for the directed case.

\begin{theorem}[See Section~\ref{sec:undirected}]\label{thm:ubUndir_intro}
    Fix $\eps\in(0,1)$. For any dataset $P\in \R^d$, we can construct an undirected graph $H$ on vertices $P\cup S$, such that for any $p,q\in P$, 
    $$
    \|p-q\|_2\le \d_H(p,q) < (1+\eps) \cdot \|p-q\|_2.
    $$
    Furthermore, the construction time, the number of edges of $H$ and the size of $S$ are bounded by $O(n^{2-\Omega(\eps^3)}\log \Delta)$, where $\Delta$ is the aspect ratio of $P$. 
\end{theorem}

\paragraph{Sub-quadratic algorithms for the Earth-Mover Distance in high dimensions problem} We highlight one concrete algorithmic application of our spanner construction. In particular, we consider the problem of Earth-Mover Distance in the high-dimensional spaces: given two $n$-points sets $A,B\subset \R^d$, compute the smallest cost bi-chromatic matching $\argmin_{\pi:A\to B} \tfrac{1}{n}\sum_{a\in A} \|a-\pi(a)\|$ where $\pi$ ranges over all permutations. In different communities, the EMD problem is also termed optimal transport, Wasserstein distance, Monge-Kantarovich, and others.\footnote{Formally, all these are defined over {\em distributions} over $n$ points, equivalently weighted pointsets. All our results apply to the weighted version as well, and hence we ignore the difference for simplicity of presentation.}

The basic problem is to compute the (near-) optimal $\pi$ efficiently. When dimension $d$ is small, a long progression of results, starting from \cite{Vai89-EMD} and culminating in \cite{khesin2019preconditioning} yielded an $(1+\eps)$-approximate algorithm running in time $\tilde O(n)\cdot \eps^{O(d)}$.

Another line of research concerns the general metric (or cost) setting, i.e., more general than the Euclidean setting. Here, the best possible runtime is $\sim n^2$, which is necessary in general. One approach uses Sinkhorn iteration algorithm to obtain additive error of $\eps \max_{p,q\in A,B} \|p-q\|$ \cite{cuturi2013sinkhorn, altschuler2017near}. A different approach uses faster linear programming algorithms to solve the standard bi-partite matching in time near-linear in the size of the (complete) bi-partite graph on $A\times B$ \cite{van2021minimum, chen2022maximum} --- this approach can obtain an exact result.

Nevertheless, the most common setting for general cost is still the high-dimensional Euclidean (or $\ell_1$) space (see, e.g., the experiments from \cite{altschuler2017near}, or this most recent paper \cite{agarwalhigher}). In this setting, one can hope for runtime $\ll n^2$, as the input size is only $O(nd)$. 

The best algorithm for the high-dimensional setting obtains runtime of $O(n^{1+O(1/c^2)})$ for $c$-approximation, for $c\ge 3.4$. This follows from using the spanner of \cite{his13} in the following general approach. One can take the pointset $A\cup B$ and build a spanner $H$ on it. Then the EMD problem reduces to solving the ``graphical EMD'' case: EMD under the metric defined  by the shortest path distance in $H$. The latter problem is equivalent to the problem of uncapacitated min-cost flow, or transshipment problem. This problem in turn can be solved in time near-linear in the graph size \cite{sherman2017generalized, andoni2020parallel, li2020faster}. While these algorithms are for undirected graphs (and hence would require an undirected spanner), the most recent algorithm does solve the min-cost flow problem in directed graph in near-linear time \cite{chen2022maximum}. 

Thus, we obtain the following statement, which is direct corollary of our directed spanner Theorem~\ref{thm:upper_bound_intro} combined with \cite{chen2022maximum}:
\begin{theorem}
    For any $\eps>0$, given two sets of points $A,B\subset \R^d$ of size $n$, we can solve the $EMD(A,B)$ problem within $(1+\eps)$ approximation in time $O(n^{2-\Omega(\eps^2)}\log\Delta)$ time.
\end{theorem}
\subsection{Our techniques}

The natural place to start is to understand why \cite{his13} does not answer our main Question~\ref{q:main}, instead obtaining a subquadratic-sized spanner only for approximation $c>3.4$. First, we note that it is enough to solve the problem for some fixed scale $r>0$: for given dataset $P$, build a spanner $H_r$ where all pairs of points at distance $\le r$ have a path of length $cr$, while no pairwise distance contracts. For this, the authors use Locality-Sensitive Hashing, originally designed to solve the Approximate Near Neighbor Search (ANNS) problem in $\R^d$:
\begin{definition}[\cite{HIM12}]
    Fix dimension $d$, approximation $c>1$, and scale $r>0$. A distribution $\cH$ over hash functions $h:\R^d\to U$, for some discrete set $U$, is called {\em $(c,p_1,p_2)$-Locality Sensitive Hashing (LSH)} if:
    \begin{itemize}
        \item for any two points $p,q\in \R^d$ with $\|p-q\|\le r$, we have $\Pr_h[h(p)=h(q)]\ge p_1$;
        \item for any two points $p,q\in \R^d$ with $\|p-q\|> cr$, we have $\Pr_h[h(p)=h(q)]< p_2$.
    \end{itemize}
\end{definition}

It is known that one can obtain $(p_1,p_2)$-LSH with $p_1\le p_2^{1/c^2+o(1)}$ for any $p_2<d^{-\omega(1)}$ \cite{AI-CACM}. The overall \cite{his13} algorithm proceeds as follows. Sample $L\approx 1/p_1$ hash functions $h$ from $(c/2,p_1,p_2)$-LSH where $p_2=1/n^3$. Each such hash function partitions the dataset $P$ into buckets. In each bucket, we connect all points to a fixed point with an edge of length $cr/2$ --- i.e., add a star to the graph $H_r$. One can immediately deduce that points at distance $\le r$ will be connected by 2-hop path of length $cr$, and that the total size is $O(nL)=O(n/p_1)=O(n^{1+12/c^2+o(1)})$. Furthermore, since $p_3=1/n^3$, one can argue that no pair at distance $<cr$ ends up in the same bucket.

Note the size is sub-quadratic only for $c>\sqrt{12}>3.4$, and there are two contributions to the exponent 12. The first one is the ``star'' gadget inside the bucket: most colliding points are connected by a 2-hop path that imposes a factor-2 stretch. The second one is the fact that we need to set $p_2=1/n^3$ to ensure that no pair of points at distance $>cr$ collide in a bucket.

As we discuss below, both these sources of stretch loss are intrinsic to the algorithm and require new ideas to overcome.

\paragraph{Lower bound.} First, we show that we cannot obtain approximation $c<2$ using vanilla spanners or even spanners with Steiner points in the Euclidean/Hamming metric. Indeed, consider a random pointset $P$ on the surface of a unit sphere in $\R^d$. Then all distances are concentrated around $\sqrt{2}$. It is immediate that, for any spanner $H_r$, any pair of points without a direct edge will have stretch 2. One can also show that even if we have Steiner points in $\R^d$, then a subquadratic-sized spanner much have stretch at least $\sqrt{2}$. For intuition, note that the best Steiner point for $P$ is the center of the sphere. A similar lower bound holds for Steiner spanners in the Hamming space.

\paragraph{Upper bound.} To deal with the first source of approximation from \cite{his13} algorithm, we consider adding Steiner vertices to the spanner that does not represent points in the metric itself. In particular, in the above equidistant case, the solution is to add a node $s$ to the graph, and connect it to each of the points in $P$ with an edge of length $cr/2$. We term this a ``star gadget'' below.

Now we can focus on the second source of stretch: that we need to set $p_2$ as low as $1/n^3$, which is a much more significant technical challenge. The natural first question would be: why not set $p_2\approx 0.1/n$ and hence $p_1=1/n^{1-\eps}$ for $c=1+\eps$? 

Consider a pair of points $p,q\in P$ at distance $\le r$ that we would like to collide in some bucket, in which case they are connected by a path of distance $cr$ (using our new star gadget from above). It is immediate to argue that $p,q$ will collide with probability $p_1$, and hence $1-o(1)$ over all $L$ hash functions. The tricky part is to ensure that this successful bucket does not also contain a pair of far points (i.e., at distance $>cr$), in which case the star gadget will shortcut the pair.

To analyze the probability of far pair of points falling into the bucket, we can consider, for example, the probability that a ``far'' point $f\in P$ at distance $\|q-f\|>cr$ collides with $q$. Standard analysis says that the expected number of far points is $\le p_2n$, and hence, by Markov's, there's, say, $\Omega(1)$ probability that no far point collides with $q$. The crucial caveat is that this probability is not independent of the event that $p,q$ collide: i.e., it may be the case that whenever $p,q$ collide, the number of colliding far points is much larger than the expectation.

We are lead to consider the ``3-point collision'' probability: $\Pr[(h(f)=h(q)) \wedge (h(q)=h(p))]$. If we were able to prove that these two events are essentially independent, we would be all set. Such property has indeed been considered, at least in \cite{AR-optimal} who used it for data-dependent LSH algorithms. They proved such desired independence by a very intricate reduction to the ``pseudo-random'' case: where $P$ is essentially random on the sphere, except for the pair of interest $p,q$. Alas, their reduction does not seem applicable here for a couple of reasons. The first is that the independence holds only in the case when far point $f$ is far from {\em both} $p$ and $q$. The second is that, while in \cite{AR-optimal} one can think of $p$ being a ``rare'' close point to $q$ (this helps in the reduction), in our setting, there could be both $\Omega(n)$ close points and $\Omega(n)$ far points for most $q$'s!

Our approach is different and uses two main ideas: the first enough to get directed spanners, and the second necessary for the undirected spanners.

\paragraph{Upper bound: first idea.}
We circumvent the ``3-point collision'' analysis by considering a different setting of the hashing, roughly corresponding to the ANNS algorithm with sub-polynomial query time. In particular, we use the hashing scheme from \cite{alrw17}. 
Think of the dataset as being two sets $P$ and $Q$, and we need to worry only about the collision of points $p\in P$ vs $q\in Q$ (in the end we take $Q=P$). Then \cite{alrw17} design a hashing scheme such that for some $L=n^{O(1)}$, there's a function $h_P:\R^d\to 2^{[L]}$ which describes for each point $p\in P$ which set of buckets it is hashed to, and there's an equivalent $h_Q:\R^d\to 2^{[L]}$ function. Then we have that, for any pair $p,q$ at distance $\le r$, we have that $h_P(p)\cap h_Q(q)\neq \emptyset$ with constant probability. Furthermore, $|h_P(p)|=n^{\rho_P+o(1)}$ and $|h_Q(q)|=n^{\rho_Q+o(1)}$ for $\rho_Q,\rho_P\ge 0$ satisfying some relation, and the runtime to compute $h_P(p), h_Q(q)$ is roughly proportional to their output size. It is possible to set $\rho_Q=0$ in which case $\rho_P\approx 1/\eps^2$ (we note that this bound is tight \cite{alrw17}).

Our solution uses the above asymmetric hashing as follows: split the dataset $P$ into $\approx n^{1-\eps^2/2}$ groups of size $m=n^{\eps^2/2}$. For each group, preprocess as above, taking a total of $\approx n/m\cdot m^{\rho_P}<n^{1.5}$ time, and then query each $q\in Q$, taking a total of $\approx n\cdot n/m=n^{2-\eps^2/2}$ time. We can then prove that, for every close pair $p\in P,q\in Q$, with probability at least $1/2$, there exists a bucket where 1) $p,q$ collide, but 2) contains no point $f$ that is far from $q$.

The remaining caveat is that we need to verify that a bucket succeeded --- all hashed $p,q$ are close to each other --- as we cannot add star gadget to the failed ones (otherwise, we create shortcuts). For this, we need to employ {\em furthest} neighbor search data structure. A na\"ive adaptation of the best currently known data structure, would result in extra runtime $\approx n^{2-\eps^3}$, a slow-down we manage to avoid.

\paragraph{Upper bound: second idea.} The above procedure obtains only a directed spanner since we guarantee distances between $p\in P$ and $q\in Q$ but among $p,p'\in P$ pairs or $q,q'\in Q$ pairs. This turns out to be much more difficult to ensure.

In fact, our original ``star gadget'' is not even enough to get an undirected spanner. In particular, consider the case when, we have $n/2$ points $P$ drawn from sphere of radius $a=2$ and $Q$ drawn sphere of radius $b=1$. Then the distance between a pair $p\in P,q\in Q$ is typically $\sqrt{a^2+b^2}=\sqrt{5}$; fix $r=\sqrt{5}$. At the same time the distance between points $p,p'\in P$ is $\sqrt{2}a=\sqrt{8}>r$. It seems impossible to construct an undirected spanner of sub-quadratic size with stretch $1+\eps$ using the ``star gadget'' from before --- because it would shortcut $P$-to-$P$ distances. 

Instead, we introduce a new gadget: {\em asymmetric star gadget}. In particular, in the above example, we have a Steiner vertex $s$ that is connected with distance $\sqrt{2}a/2$ to points $p\in P$ and with distance $\sqrt{a^2+b^2}-\sqrt{2}a/2\ge\sqrt{2}b/2$ to points $q\in Q$. Now the distance $P$-to-$Q$ is $r$; $P$-to-$P$ is $\sqrt{2}a$; and $Q$-to-$Q$ is $\ge\sqrt{2}b$ as well. While the $Q$-to-$Q$ distance is stretched (it will be taken care of at a different scale $r$), the $r$-scale distances are taken care of, and there are not shortcuts created.

To be able to use this gadget, we start from where we left off with the directed spanner: a pair of sets of $P'\subset P, Q'\subset Q$ such that each $p\in P'$ is close to each $q\in Q'$. We show that in this case, there exists $a,b$ with $a+b\le r$, such that we can decompose $P'$ (and $Q'$) into smaller parts so that the diameter of each part is $(1+\eps)a$ (and $(1+\eps)b$ respectively). The number of such parts is bounded by $P'^{1-\Theta(\eps)}$ and $Q'^{1-\Theta(\eps)}$ respectively. Then we use the asymmetric gadget on each pair of such parts.

\subsection*{Acknowledgements}

We thank Negev Shekel Nosatzki for important discussions and suggestions for the undirected spanner construction in Section \ref{sec:undirected}. Research supported in part by NSF (CCF2008733) and ONR (N00014-22-1-2713).

\section{Lower Bound for Spanners, with or without Steiner Points}
\label{sec:lowerBound}

In this section, we prove the lower bounds on the size of the spanner for geometric pointsets. We consider spanners that contains Steiner nodes corresponding to points in the metric. We note that since the spanner cannot shortcut any pair, for any spanner edge connecting two points $x,y\in \R^d$, the optimal length of the edge is the distance between points $x$ and $y$.

We prove lower bounds for both the Hamming and Euclidean ($\ell_2$) spaces separately.\footnote{While it is known that $\ell_2$ can be embedded into $\ell_1$ (and $\ell_1$ into $\ell_2$-squared), a lower bound for the $\ell_2$ ($\ell_1$) space does not immediately imply a lower bound for the $\ell_1$ (resp. $\ell_2$) space since the Steiner points are to be taken in the considered space.}
Both lower bounds use the following lemma, which we prove in Section~\ref{sec:pathsLemma}.

\begin{lemma}\label{lem:lower_bound_path}
Consider a graph $G= (V,E)$ and integer $k$. Let $P_x,P_y\subseteq V$ be any disjoint sets. Suppose $|P_x| = |P_y| = n$ and $|V| \leq |E| = o(n^2/k^2)$. For every $x\in P_x$ and $y\in P_y$, let $S_{x,y}$ be any path from $x$ to $y$ in $G$.
Then, there exists $k$ distinct points $x_1,\cdots,x_k\in P_x$ and $k$ distinct points $y_1,\cdots,y_k\in P_y$, such that 
there is a node that lays on all paths $S_{x_1,y_1}, S_{x_2,y_2}, \cdots, S_{x_k,y_k}$.
\end{lemma}

\subsection{Lower bound in the Hamming metric}\label{sec:lb_hamming}
We prove the following lower bound on Steiner spanner size in the $\ell_1$ metric.
\begin{theorem}[Restatement of Theorem~\ref{thm:lower_bound_intro}]\label{thm:lower_bound}
For any $\eps \in (0,1)$, there exists a point set $P\subset \{0,1\}^d$ such that the following holds.
Let $H = (P\cup S, E)$ be any spanner that can use extra Steiner points $S\subseteq \{0,1\}^d$, but can only use the $\ell_1$ distance as the edge weight between two points.
Suppose for all $u,v\in P$, the distance between $u$ and $v$ on graph $H$ satisfies
\[
\|u-v\|_1 \leq \d_H(u,v) \leq (1+\eps) \|u-v\|_1,
\]
then $H$ must use $\Omega(n^2(1-\eps)^4)$ edges.
\end{theorem}
\begin{proof}[Proof of Theorem~\ref{thm:lower_bound}]
Let $P$ be chosen as every point is i.i.d. uniformly sampled from the hamming cube $\{0,1\}^d$ with $d=\omega(\log^2 n)$. With high probability, every pair of points in $P$ has distance $\in d/2 \pm o(1)d$. 

For every two points $x,y \in \{0,1\}^d$, we define $A_{x,y} := \{ z\in \{0,1\}^d \mid \|x-z\|_1 + \|y-z\|_1 \leq \frac{\delta}{2}\cdot d + \|x-y\|_1 \}$, where $\delta = (1+\eps)/2$ is taking average of $1$ and $\eps$.
Intuitively, for $x,y\in P$, the set $A_{x,y}$ contains all possible Steiner points $s$ in $\{0,1\}^d$ such that the path $x\rightarrow s\rightarrow y$ won't have large approximation: if $\|x-s\|_1 + \|s-y\|_1 \leq (1+\eps)\|x-y\|_1$, then $s\in A_{x,y}$ since $(1+\eps)\|x-y\|_1 \leq \frac{\delta}{2}\cdot d + \|x-y\|_1$.

The following claim shows that $A_{x,y}$ has small volume.

\begin{claim}\label{cla:pr_a_x_y}
For every $p\in \{0,1\}^d$, if $x,y$ are uniformly sampled from $\{0,1\}^d$, then
\begin{align*}
\Pr_{x,y}[p \in A_{x,y}] \leq \exp(-\frac{1}{8}(1-\delta)^2d).
\end{align*}
\end{claim}
\begin{proof}
Define $S = \{i\in [d] \mid x_i=y_i\}$ be the set of coordinates where $x$ and $y$ agree on. $A_{x,y}$ has an equivalent definition  $A_{x,y} = \{z \in \{0,1\}^d \mid \|z_S - x_S\|_1 \leq \frac{\delta}{4} \cdot d\}$. For $i\in[d]$, let $X_i$ be the random variable that $X_i = 1$ if $i\in S$ and $p_i = x_i$, otherwise $X_i = 0$. 
If $x,y$ are sampled uniformly randomly from $\{0,1\}^d$, then each $X_i = 1$ with $\frac{1}{4}$ probability and $X_i = 0$ with $\frac{3}{4}$ probability, and all $X_i$ are independent. And $p\in A_{x,y}$ is equivalent to $\sum X_i \leq \frac{\delta}{4}\cdot d$.

By Chernoff bound $\Pr[X\leq (1-\alpha)\E[X]]\leq e^{-\alpha^2\E[X]/2}$, replacing $\alpha$ with $1-\delta$, we have 
\begin{align*}
    \Pr_{x,y}[p \in A_{x,y}] = \Pr[\sum_i X_i \leq \frac{\delta}{4}\cdot d] \leq \exp(-\frac{1}{8}(1-\delta)^2d).
\end{align*}
\end{proof}

We are now ready to proceed with the main argument. For any $P$, let $H$ be any spanner stated in this theorem. Suppose $H$ has $o(n^2/k^2)$ edges for some $k$ tbd. By Lemma~\ref{lem:lower_bound_path}, there must exist $2k$ distinct points $x_1,y_1,\cdots,x_k,y_k \in P$ such that the shortest path between $(x_1,y_1),\cdots (x_k,y_k)$ intersect on some vertex $v\in H$ (which could be Steiner or an original point in $P$). Since the edge weight can only use the actual Hamming distance, we must have $\|x_i-v\|_1+\|y_i-v\|_1\leq (1+\eps)\|x_i-y_i\|_1 \leq \|x_i-y_i\|_1 + \frac{\delta}{2} d$ for all $i$. Thus, $v$ must be in every $A_{x_i,y_i}$ by definition of $A_{x,y}$.

Define the event $E$ as follows: there exists a point $p\in \{0,1\}^d$, and $2k$ distinct points $x_1,y_1,\cdots,x_k,y_k\in P$ such that $p\in A_{x_i,y_i},~\forall i\in[k]$. We have argued that if the claimed spanner with $o(n^2/k)$ edges exists for every $P$, then $\Pr_P[E] = 1$.
We will show next that when $k > 24(1-\delta)^{-2}$, the probability $E$ happens (over the random choice of $P$) is less than $1$. Hence, there exists a dataset $P^*$ for which $E$ does not happen, and therefore a claimed spanner $H$ does not exist for $P^*$.

By Claim~\ref{cla:pr_a_x_y}, for a fixed $p\in \{0,1\}^d$, we have
\begin{align*}
\Pr_{x,y\sim \{0,1\}^d}[p \in A_{x,y}] \leq \exp(-\frac{1}{8}(1-\delta)^2d)
\end{align*}

Therefore, for $k> 24(1-\delta)^{-2}$,
\begin{align*}
    & ~ \Pr[\exists \text{~distinct~}x_1,y_1,\cdots,x_k,y_k\in P,~\mathrm{such~that}~p \in \cap_{i\in [k]}A_{x_i,y_i}] \\
    \leq & ~  n^{2k}\cdot (\Pr_{x,y}[p \in A_{x,y}])^k \\
    \le & ~ \exp(2k\ln n-\frac{k}{8}(1-\delta)^2d)\\
    \leq & ~  \exp(-2d).
\end{align*}

Thus, we conclude $\Pr[E] < 2^d\cdot \exp(-2d)=o(1)$ by union bound on all points $p\in\{0,1\}^d$, and therefore the theorem follows.
\end{proof}

\subsection{Lower bound in $\ell_2$ metric}\label{sec:lb_euclidean}

We now prove the lower bound for the $\ell_2$ metric:

\begin{theorem}[$\ell_2$ lower bound]\label{thm:lower_bound_l2}
For any $\eps \in (0,\sqrt{2}-1)$, there exists a dataset $P\subset S^{d-1}(0,1)$, such that the following holds.
Let $H = (P\cup S, E)$ be any spanner that can use extra Steiner points $S\subset \R^d$, but can only use the $\ell_2$ distance as the edge weight between two points.
Suppose for all $u,v\in P$, the distance between $u$ and $v$ on graph $H$ satisfies
\[
\|u-v\|_2 \leq \d_H(u,v) \leq (\sqrt{2}-\eps) \|u-v\|_2,
\]
then $H$ must use $\Omega(n^2\eps^4/\log^2 n)$ edges, where $n=|P|$.
\end{theorem}

First, we give some definitions. For any $\eps\in (0,\sqrt{2}-1)$ and two points $x,y\in \R^d$, let $A_{x,y}^{\eps}$ be defined as 
$A_{x,y}^{\eps} := \{ p\in \R^d \mid \|x-p\|_2 + \|y-p\|_2 \leq (\sqrt{2}-\eps) \|x-y\|_2 \}$. If $\eps$ is clear from the context, we write it as $A_{x,y}$.

The following lemma shows that for a fixed point $p$, it is rare $A_{x,y}$ contains $p$ for a random pair of points $x$ and $y$.
\begin{lemma}\label{lem:p_a_x_y_l2}
For $\eps \in (0,\sqrt{2}-1)$, let $p \in B^d(0,1)$ be any point in the unit ball. Let $x,y$ be sampled uniformly from unit sphere $S^{d-1}$. Then
\[
\Pr_{x,y\sim S^{d-1}}[p \in A_{x,y}^{\eps} \mid \|x-y\|_2\in (1\pm o_{\eps}(1))\sqrt{2}] \leq e^{-\eps^2d/2}.
\]
\end{lemma}
\begin{proof}
It is enough to bound $\Pr_{x,y\sim S^{d-1}}[p \in A_{x,y} \mid \| x-y\|_2 = 2t]$ for any fixed $t\in (1\pm o_{\eps}(1))\sqrt{2}/2$. By symmetry, this is equivalent to 
\[
\Pr_{p'\sim S^{d-1}(0,r)}[ p'\in A_{x,y}] = \frac{\Vol(A_{x,y} \cap S^{d-1}(0,r))}{\Vol(S^{d-1}(0,r))},
\]
where $r := \|p\|_2$, and $x,y$ can be any two fixed points on $S^{d-1}$ that $\|x-y\|_2 = 2t$, say, $x = (-t,\sqrt{1-t^2},0_{d-2}),y = (t,\sqrt{1-t^2},0_{d-2})$. 

Note that $A_{x,y}$ is an ellipsoid with focal points on $x$ and $y$. Let $c=(0,1,0_{d-2})$. Let 
\begin{align*}
s := & ~  \min_{z\in A_{x,y}} \langle z,c\rangle \geq \sqrt{1-t^2} - \sqrt{ (\sqrt{2}-\eps)^2t^2 - t^2}\\
= & ~ \frac{1}{\sqrt{2}} + o_{\eps}(1) - \frac{\sqrt{1-2\sqrt{2}\eps + \eps^2}}{\sqrt{2}}-o_{\eps}(1)\\
\geq & ~ \eps.
\end{align*}

This means $A_{x,y} \cap S^{d-1}(0,r) \subseteq \textrm{Cap}(r,c,s)$, where $\textrm{Cap}(r,c,s)$ is the sphere cap including all points $x\in S^{d-1}(0,r)$ for which $x\cdot c\geq s$. Using the bound on the volume of a high dimensional sphere cap given by \cite{tkocz2012upper}, we get
\[
\frac{\Vol(\textrm{Cap}(r,c,s))}{\Vol(S^{d-1}(0,r))} \leq e^{-d (s/r)^2/2} = e^{-\eps^2d/2}.
\]

\end{proof}

\begin{proof}[Proof of Theorem~\ref{thm:lower_bound_l2}]
Let $P$ be generated as every point is i.i.d. uniformly sampled from unit sphere $S^{d-1}(0,1)$ with $d=\omega(\log^2 n)$. With high probability, every pair of points in $P$ has distance $\in \sqrt{2} \pm o(1)$.

Let $H$ be any spanner stated in this theorem with $o(n^2/k^2)$ edges for $k=10\log n/\eps^2$. Wlog, we can assume every Steiner point of $H$ is in the unit ball $B^d(0,1)$, since otherwise we can project them onto the unit sphere.
By Lemma~\ref{lem:lower_bound_path}, there must exists $2k$ distinct points $x_1,y_1,\cdots, x_k,y_k\in P$ such that the shortest path between $(x_1,y_1),\cdots,(x_k,y_k)$ intersect on some point $v\in H$. 
Since the edge weight can only use the actual distance in the Euclidean space, we must have $\|x_i-v\|_2+\|y_i-v\|_2\leq (\sqrt{2}-\eps)\|x_i-y_i\|_2$ for all $i$. Let $v'$ be the closest point to $v$ such that every coordinate of $v'$ is a multiple of $\frac{\eps}{10\sqrt{d}}$, and thus $\|v-v'\|_2\leq \eps/10$, and therefore, $v'$ satisfies that for all $i$,
\[
\|x_i-v'\|_2 + \|y_i-v'\|_2\leq (\sqrt{2}-\eps/2)\|x_i-y_i\|_2.
\]

Let $N$ include all points in ball $B^{d}(0,r)$ whose coordinates are multiple of $\frac{\eps}{10\sqrt{d}}$ and let the event $E$ be: there exists a point $p\in N$ and $2k$ distinct points $x_1,y_1,\cdots x_k,y_k\in P$ such that for all $i\in [k]$ $p\in A^{\eps/2}_{x_i,y_i}$. The above argument indicates that if such a spanner exists, then $E$ must happen. Next, we will show that $E$ happens with probability $<1$ for a random dataset $P$; therefore, we assert that such spanner $H$ does not exist for some $P$.

Let $p\in N$ be any point. Since every pair of points in $P$ has distance $\in \sqrt{2} \pm o(1)$, by Lemma~\ref{lem:p_a_x_y_l2}, we have
\begin{align*}
& ~ \Pr[\exists x_1,y_1,\cdots x_k,y_k,~~\text{s.t.}~~p\in \cap_{i} A^{\eps/2}_{x_i,y_i}]\\
\leq & ~ n^{2k}\cdot \left(\Pr_{x,y}[p\in A_{x,y} \mid \|x-y\|_2=\sqrt{2}\pm o(1) ]\right)^k\\
\leq & ~ \exp(2k\ln n-\eps^2 d k/8)\\
= & ~ \exp(-d\log n).
\end{align*}

By union bound on all $p\in N$, we have
\[
\Pr[E]\leq \exp(-d\log n)\cdot (20\sqrt{d}/\eps)^d = o(1),
\]
therefore, the theorem follows.
\end{proof}

\subsection{Proof of Lemma~\ref{lem:lower_bound_path}}
\label{sec:pathsLemma}

\begin{proof}[Proof of Lemma~\ref{lem:lower_bound_path}]
Suppose there is a graph $G$ for which the property doesn't hold. We will turn $G$ into another bipartite graph $H = (P_x \cup P_y, E')$ with less than $n^2$ edges but can connect all pairs of points $x\in P_x$ and $y\in P_y$ by a direct edge. Thus, by proof of contradiction, we can conclude such graph $G$ doesn't exist.

First of all, we assume every path $S_{x,y}$ never touches $P_x$ or $P_y$ except its endpoints, since otherwise, we can create dummy points for every point in $P_x$ and $P_y$ and let the path $S_{x,y}$ use dummy points. If the new graph has $k$ paths intersecting on one point, then so does the original graph.

Next, to simplify the proof, we give direction to edges. Let $S_{x,y}$ be any path, and let $\{v_1,\cdots,v_k\}$ be points on the path in order. We say the directed edge $e=(a,b)\in S_{x,y}$ if $a=v_{i}$ and $b=v_{i+1}$ for some $i$. We assume $G$ only has useful edges. From now on, $G$ is a directed graph and
\[
E = \bigcup_{x\in P_x,y\in P_y} \bigcup_{(a,b)\in S_{x,y}} \{(a,b)\}.
\]

To turn $G$ into $H$, we have three intermediate steps -- from $G$ to $G_1$, $G_2$, $G_3$, then $H$, each graph has better properties. In the first graph $G_1 = (V_1, E_1)$, we will have $|V_1| = O(|V|)$, $|E_1| = O(|E|)$, and every vertex is \textit{light} which we will explain later. The procedure of making $G_1$ is as follows.

For every vertex $v\in V$, let $U(v)$ be initially empty. For every path $S_{x,y}$ with $x\in P_x$ and $y\in P_y$, let $\{v_1,\cdots,v_k\}$ be the points on the path. For every $i\in [k]$, we add $(x,y)$ to $U(v_i)$. 

We show that we modify the graph to obtain a simpler condition --- that $U(v)$ contains $<k$ of either distinct $x$'s or distinct $y$'s. 
For every vertex $v\in V\backslash (P_x\cup P_y)$, $U(v)$ can be regarded as a bipartite graph with a collection of edges between $P_x$ and $P_y$. In this bipartite graph, the maximum matching is less than $k$, otherwise there are $k$ paths with distinct endpoints intersecting on $v$. 

\begin{theorem}[Extended Hall’s theorem]
For any bipartite graph $G=(L,R,E)$ with $|L| = |R| = n$, the cardinality of maximum matching is
\[
\mu(G) = n-\max_T ( |T| - |N(T)| ),
\]
where $T$ range over all subset of $L$ and $R$ separately, and $N(T)$ denote the set of neighbors of the set $T$ in $G$.
\end{theorem}

Apply the above theorem on the bipartite graph $U(v)$, there must be a set $T\subseteq P_x$ or  $T\subseteq P_y$ such that $n - |T| + |N(T)| = \mu(U(v)) < k$. We must have $|T| \geq n-k$ and $|N(T)| \leq k$. The analysis for the cases $T\subseteq P_x$ and $T\subseteq P_y$ would be similar, so we assume the first case. Let $T_L = T$ and $T_R = P_y \backslash N(T)$. Note that there is no path $S_{x,y}$ in $G$ crossing $v$ with $x\in T_L$ and $y\in T_R$. In the new graph $G_1$, we create three copies of $v$: $v_{T_L, \bar{T}_R}$ and $v_{\bar{T}_L, T_R}$ and $v_{\bar{T}_L, \bar{T}_R}$, where $\bar{T}_L := P_x \backslash T_L$ and $\bar{T}_R := P_y \backslash T_R$. So $|V_1| \leq 3 |V|$. For each edge $(u,v)$ in $G$, we create at most $9$ edges linking each pair of copies of $u$ and $v$. Thus, $|E_1| \leq 9 |E|$.

Let $\{v_1,\cdots,v_l\}$ be vertices on path $S_{x,y}$. For all $2\leq i \leq k-1$, we replace $v_i$ with its copy $v_{S,T}$ where $x\in S,y\in T$ and get a path $S'_{x,y}$ in $G_1$.
In the new graph $G_1$, for every vertex $v\in V_1$, let $L(v),R(v)$ be initially empty. For every path $S'_{x,y}$ with $x\in P_x$ and $y\in P_y$, let $\{v_1,\cdots,v_l\}$ be the points on the path. For every $i\in [l]$, we add $x$ to $L(v_i)$ and add $y$ to $R(v_i)$.

Since $|\bar T_L|, |\bar T_R| \leq k$ from the construction of $G_1$, for every vertex $v\in V_1\backslash (P_x\cup P_y)$, we have either $|L(v)|\leq k$ or $|R(v)|\leq k$. If $|L(v)|\leq k$, we call $v$ \textit{left-light}, otherwise we call $v$ \textit{right-light}. So there are four types of points in $G_1$: $P_x$, $P_y$, \textit{left-light} and \textit{right-light} points.

For any graph $G=(V,E)$ and a vertex $v\in V$, we call $\textrm{in}_G(v)$ as the set of edges entering $v$, and $\textrm{out}_G(v)$ as the set of edges leaving $v$. In some context, we also refer $\textrm{in}_G(v)$ $\textrm{out}_G(v)$ as those vertices connected with an edge entering/leaving $v$.

Then we construct the next graph $G_2 = (V_2,E_2)$ as follows. Initially, $E_2 = E_1$. Then, for every \textit{left-light} vertex $v$, we delete all edges in $\textrm{in}_{G_1}(v)$ and add edge $(a,v)$ for every $a\in L(v)$. For every \textit{right-light} vertex $v$, we delete all edges in $\textrm{out}_{G_1}(v)$ and add edge $(v,b)$ for every $b\in R(v)$. Intuitively, the construction is replacing edges with shortcuts. One can check that every pair of points $x\in P_x,y\in P_y$ is still connected with a path in $G_2$. The number of edges is bounded by $|E_2| \leq |E_1| + |V_1|\cdot k$. The purpose of constructing $G_2$ is to have $\textrm{in}_{G_2}(v) \subseteq P_x$ for \textit{left-light} vertices $v$, and $\textrm{out}_{G_2}(v) \subseteq P_y$ for \textit{right-light} vertices $v$. Thus, $G_2$ is in fact a four-layered graph ordered from $P_x$ -- \textit{left-light} vertices-- \textit{right-light} vertices -- $P_y$, and every edge must go strictly from lower level to higher level.

To construct $H$ from $G_2$, we need to remove all points in $V_2 \backslash (P_x\cup P_y)$. Define the removing procedure applied on vertex $v$ on graph $G^*$, $\textsc{Remove}_{G^*}(v)$ as follows: delete all edges connecting $v$, and then for every vertex $a\in \textrm{in}_{G^*}(v)$ and $b\in \textrm{out}_{G^*}(v)$, add an edge from $a$ to $b$.

We apply $\textsc{Remove}_{G_2}(v)$ on every \textit{left-light} vertices $v$ in $G_2$ to get graph $G_3=(V_3,E_3)$, then apply $\textsc{Remove}_{G_3}(v)$ on every \textit{right-light} vertices $v$ in $G_3$ to get graph $G_4=(P_x\cup P_y,E_4)$. The order of deleting $v$ does not affect the outcome since there are no edges connecting two \textit{left-light} vertices, nor two \textit{right-light} vertices. We have
\begin{align*}
|E_3| \leq |E_2| + \sum_{v~is~\textit{left-light}}k\cdot \textrm{out}_{G_2}(v) \leq |E_2| + k|E_1|,
\end{align*}
where the last step follows by $\textrm{out}_{G_2}(v) =\textrm{out}_{G_1}(v)$ for \textit{left-light} vertices $v$.

Furthermore, we have
\[
|E_4| \leq |E_3| + \sum_{v~is~\textit{right-light}}k\cdot \textrm{in}_{G_3}(v) \leq (k+1)|E_3|.
\]

Thus, $|E_4| \leq (k+1)|E_3| \leq (k+1)k (|V_1| + |E_1|) =O(k^2)\cdot |E| < n^2$. And $G_4$ is our $H$ -- a bipartite graph that connects all pairs of points from $x\in P_x$ and $y\in P_y$ with less than $n^2$ edges.
\end{proof}

\section{Directed Non-metric Steiner Spanners}
\label{sec:directedSpanner}
In this section we construct a directed non-metric Steiner spanner, proving Theorem~\ref{thm:upper_bound_intro}.

\begin{theorem}[Restatement of Theorem~\ref{thm:upper_bound_intro}]\label{thm:directed_spanner}
Fix $\eps\in(0,1)$. For any dataset $P,Q\in \R^d$, there exists a directed graph $H$ on vertices $P\cup S$, such that for every $p,q\in P$, 
    $$
    \|p-q\|_2\le \vec{\d}_H(p,q) < (1+\eps) \cdot \|p-q\|_2.
    $$
Furthermore, the number of edges of $H$, the size of $S$, and the construction time are all bounded by $O(n^{2-\Omega(\eps^2)}\log\Delta)$, where $\Delta$ is the aspect ratio of $P$.
\end{theorem}

We prove this theorem by reducing to the one-scale case: we fix distance $r$, and we construct a spanner $H_r$ such that, for all pairs of points within distance $[r, (1+\eps)r]$, they are all connected in $H_r$ with a proper path. Here is the formal statement.
\begin{lemma}\label{lem:core}
Let $P\subset \R^d$ be a point set on the unit sphere. Let $n = |P|$, $r = (\log n)^{-1/8}$, and $\eps$ is the approximation.
There is an algorithm running in $n^{2-\Omega(\eps^2)}$ time, that can generate a directed graph $H = (P \cup S, E)$ with $|E| \leq n^{2-\Omega(\eps^2)}$. Here $S$ is the set of extra non-metric Steiner points. With high probability, $H$ satisfies:
\begin{enumerate}
\item[(1)] $\forall p,q\in P$, $\vec{\d}_H(p,q) \geq \|p-q\|_2$;
\item[(2)] $\forall p,q\in P$ with $\|p-q\|_2 \leq r$, there is a 2-hop directed path in $H$ from $p$ to $q$ of length $\le (1+\eps)r$.
\end{enumerate}
\end{lemma}

The proof of this lemma is in the rest of this section. For now, we finish the proof of our main theorem using Lemma~\ref{lem:core}.

\begin{proof}[Proof of Theorem 3.1]
In Lemma~\ref{lem:core}, we show how to build a spanner, so that every pair of points $p,q\in P$ with $\|p-q\|_2 \in [r,(1+\eps)r]$ are handled, given any parameter $r$.
We can do doubling on $r$ from least distance to largest distance to create a sequence of spanners $H_1,\cdots,H_{\log \Delta}$. Then $H=\cup H_i$ is the final spanner.

There is only one issue to solve in applying Lemma~\ref{lem:core}: that we need $P$ to be on the sphere and $r$ has to be about $1/\log^{1/8} n$. Indeed, it is possible to reduce any Euclidean instance to such an instance, while preserving the distances at the scale $r$. See one such reduction in Corollary A.2 from \cite{privateNNS23} (based on \cite[Lemma 6]{bartal2011dimreduction}).

\end{proof}

\subsection{Algorithm for one scale}
The algorithm heavily uses the data-independent (asymmetric) locality sensitive hashing from \cite{alrw17} as a subroutine. We use the following primitive from \cite{alrw17}; while this lemma is not explicitly given in \cite{alrw17}, we show how it is obtained in Appendix~\ref{sec:lsh}.

\begin{lemma}[Data-independent LSH from Section~3 of \cite{alrw17}]\label{lem:lsh}
Let $r = (\log n)^{-1/8}$ be a fixed parameter, $c=1+\eps$ be the approximation ratio. There is a data structure that, takes a random seed $s$, implicitly \footnote{$T$ can be super-polynomially large.} gives two random collections of subspaces $A_1,A_2,\cdots,A_T\subset \R^d$ and $B_1,B_2,\cdots, B_T \subset \R^d$. For any point $p$, denote $\mathcal{A}_p = \{A_i : p \in A_i\}$ as the set of subspaces $A_i$ that contains $p$. Similarly, define $\mathcal{B}_p= \{B_i : p\in B_i\}$.
Then, for any $n=\omega(1)$, we have
\begin{enumerate}
\item $\forall p\in S^{d-1}$, $\E_s[|\mathcal{A}_p|] \leq n^{1/\eps^2 + O(1/\eps)}$;
\item $\forall p\in S^{d-1}$, $\E_s[|\mathcal{B}_p|] = n^{o(1)}$;
\item $\forall p,q\in S^{d-1}$ that $\|p-q\|_2 \leq r$, $\forall i\in [T]$,
\[
\Pr_s[p\in A_i \mid q\in B_i] \geq n^{-o(1)};
\]
\item $\forall p,q\in S^{d-1}$ that $\|p-q\|_2 \geq c\cdot r$, $\forall i\in [T]$,
\[
\Pr_s[p\in A_i \mid q\in B_i] \leq \frac{1}{n};
\]
\item $\forall p\in S^{d-1}$, one can find all $A_i \in \mathcal{A}_p$ in $O(|\mathcal{A}_p|\cdot n^{o(1)}$ time, and all $B_i \in \mathcal{B}_p$ in $O(|\mathcal{B}_p|\cdot n^{o(1)})$ time.
\end{enumerate}
\end{lemma}

\noindent \textbf{Description of algorithm for directed spanner construction.} The formal algorithm is presented as Algorithm~\ref{alg:directed}.
The algorithm takes as input a dataset $P\subseteq S^{d-1}$, a distance parameter $r\in R$, and a approximation factor $\eps$, and output a Steiner spanner graph $H$. It never create ``shortcut'', i.e., $\vec{\d}_H(p,q)$ cannot be smaller than $\|p-q\|_2$. For those $\|p-q\|_2\in [r,(1+\eps)r]$, it guarantees that $\vec{\d}_H(p,q) \leq (1+\eps)\|p-q\|_2$.

The algorithm consists of $n^{o(1)}$ rounds. In each round, for every pair of points $p,q$ at distance $\leq r$, they have $n^{-o(1)}$ chance being connected by a 2-hop path of total length of $(1+\eps)r$. Furthermore, we avoid creating a short-cut on any pair of points.
Thus, after $n^{o(1)}$ rounds, we expect our guarantee is satisfied.

In each round, we first arbitrarily partition all points into groups of size $m= n^{\eps^2/4}$. Denote groups as $P_1,\cdots,P_{n/m}$. Then, we use Locality Sensitive Hashing from Lemma~\ref{lem:lsh} (with parameters $c_{1} \leftarrow 1+\eps$ be the approximation and $n\leftarrow K m$ for some $K = |P|^{o(1)}$). Note that we use the same LSH function for all groups. 
Assume $\{A_i,B_i\}_{i\in[T]}$ are half-spaces generated by LSH function. Define sets $Q_j\leftarrow P\cap B_j$ and $P_{i,j}\leftarrow P_i \cap B_j$.  Intuitively, every point in $Q_j$ will be close to every point in $P_{i,j}$ for all $i$, because it's exactly the building and querying procedure of a nearest neighbor data structure -- in the pre-processing phase, $p\in P_{i,j}$ falls into $j$-th bucket, and in the query phase, $q\in Q_j$ queries the $j$-th bucket.

To guarantee that some $q\in Q_j$ is at distance $\leq (1+\eps)r$ from all the points in $P_{i,j}$ (which may fail to happen with non-trivial probability), we perform a {\em Furthest Neighbor Search (FNS)} check. 
For the latter, we use an FNS data structure, stated in the lemma below. It follows immediately from the reduction from \cite{I-thesis} to a nearest neighbor search data structure, which we instantiate with the data structure from~\cite{alrw17}.
\begin{lemma}[FNS data structure \cite{I-thesis, alrw17}]\label{lem:fns}
There is a $c$-approximate Furthest Neighbor Search data structure with $O(n^{1+\rho_P+o(1)})$ preprocessing time and $O(n^{\rho_Q+o(1)})$ query time, for any $\rho_Q, \rho_P>0$ satisfying the trade-off of:
\[
c^2\sqrt{\rho_Q} + (c^2-1)\sqrt{\rho_P} = \sqrt{2c^2-1}.
\]
\end{lemma}
We build an FNS data structure (Lemma~\ref{lem:fns}) for each $P_{i,j}$ with approximation $c_{2}\leftarrow 1+\eps$ and parameters $\rho_P \leftarrow 2/\eps^2$ and $\rho_Q\leftarrow 0$. For every point $q\in Q_j$, we check if every point in $P_{i,j}$ is $c_2$ approximately $(1+\eps)r$-close to $q$. If so, we can safely put $q$ into $V_{i,j}$. As a result, every point $q\in V_{i,j}$ and every point $p\in P_{i,j}$ has Euclidean distance $\leq (1+3\eps)r$ so we can create an extra Steiner node as the center and add a star gadget. We will prove later that if $p,q$ are close, they will fall into $P_{i,j}$ and $V_{i,j}$ with $n^{-o(1)}$ probability, for some $i,j$.

\begin{algorithm}[h]\caption{}\label{alg:directed}
\begin{algorithmic}[1]
\Function{Build Spanner}{$P\subseteq S^{d-1}$, $r,\eps > 0$}
\State Let $n=|P|$ and $m = n^{\eps^2/4}$
\State Initialize the spanner $H=(P,\emptyset)$ as a graph with vertex set $P$ but no edges
\For{$\textrm{it} = 1,2,\cdots,n^{o(1)}$} \Comment{Repeat $n^{o(1)}$ rounds to boost success probability}
    \State Arbitrarily partition $P$ into $n/m$ groups $P_1,\cdots P_{n/m}$ each of size $m$
    \State Let $\{A_j,B_j\}_{j\in [T]}$ be the LSH function stated in Lemma~\ref{lem:lsh}
    \State \Comment{Plug in parameter $c_1\leftarrow 1+\eps$ and $n\leftarrow Km$ for some $K=|P|^{o(1)}$}
    \State Compute $Q_j \leftarrow P\cap B_j$, $\forall j\in [T]$
    \For{$i\in [n/m]$,~$j\in[T]$}
        \State Calculate $P_{i,j} \leftarrow P_i \cap A_j$
        \State $V_{i,j}\leftarrow \emptyset$
        \State Build FNS data structure for $P_{i,j}$ using Lemma~\ref{lem:fns}
        \State \Comment{Plug in parameter $c_2\leftarrow 1+\eps$, $\rho_P\leftarrow 2/\eps^2$ and $\rho_Q\leftarrow 0$.  \label{line:alg1_fns}}
        \For{$q\in Q_j$}
            \State Query $q$ into the FNS data structure for $P_{ij}$ to get  $f$, the $c_2$-approximate furthest point
            \If{$\|q-f\|_2 \leq c_1r$} \label{line:fns_test}
                \State Add $q$ to $V_{i,j}$
            \EndIf
        \EndFor
        \State Create a new Steiner node $s_{ij}$ in $H$ \label{line:create_star_start}
        \State For all $q\in V_{i,j}$, add a directed edge $(q,s_{ij})$ with length $0$ to $H$ \label{line:add_edge_1}
        \State For all $p\in P_{i,j}$, add a directed edge $(s_{ij},p)$ with length $(1+3\eps)r$ to $H$ \label{line:add_edge_2}
        \label{line:create_star_end}
    \EndFor
\EndFor
\State \Return $H$
\EndFunction
\end{algorithmic}
\end{algorithm}

\subsection{One scale correctness: Proof of Lemma~\ref{lem:core}} \label{sec:lem_core}

We now prove Lemma~\ref{lem:core}.
We first analyze the size of sets $P_{i,j}$ and $Q_j$, and then bound the size of spanner.

\begin{observation}\label{obs:alg1}
Let $P_{i,j}$ and $Q_j$ be defined in Algorithm~\ref{alg:directed}. We have
\begin{enumerate}
\item for all $i,j$, $P_{i,j} \subseteq P_i$ and $|P_i|=m$.
\item $\sum_j |Q_j| = n^{1+o(1)}$;
\item $\sum_{i,j} \E[|P_{i,j}|] = n^{5/4+O(\eps)}$;
\item $\forall j, \sum_{i}|P_{i,j}| \leq n$, 
\item $i\in [n/m]$ and $m = n^{\eps^2/4}$.
\end{enumerate}
\end{observation}
\begin{proof}
We need to prove (2), (3), (4).

For (2), (3), by Lemma~\ref{lem:lsh}, each $q\in P$ is mapped to $(Km)^{o(1)} = n^{o(1)}$ different $Q_j$, and each $p\in P_i$ is mapped to  $(Km)^{1/\eps^2 + O(1/\eps)} = n^{1/4+O(\eps)}$ different $P_{i,j}$. For (4), we note that for every fixed $j$, any $p\in P$ can appear in  $P_{i,j}$ for at most one value of $i$.
\end{proof}

\begin{proof}[Proof of Lemma~\ref{lem:core}]
We are going to prove the lemma in four aspects: the graph our algorithm output satisfies both property (1) and (2), also the edge number and running time are both upper bounded by $n^{2-\Omega(\eps^2)}$. For simplicity, we will run Algorithm~\ref{alg:directed} with $\eps = \frac{1}{3}\eps_{\mathrm{input}}$. The lemma then follows with a substitution.

\textbf{Property 1:} $\forall p,q\in P$, $\vec{\d}_H(p,q) \geq \|p-q\|_2$.

The only situation we add edges is when creating paths between $P_{i,j}$ and $V_{i,j}$, and all paths have length $(1+3\eps)r$. Because every $q$ enters $V_{i,j}$ only after it passes FNS test (Line~\ref{line:fns_test}), which means every point in $P_{i,j}$ is $c_1c_2r < (1+3\eps)r$ close to $q$. So property (1) is true.

\textbf{Property 2:} $\forall p,q\in P$ with $\|p-q\|_2 \leq r$, there is a 2-hop directed path in $H$ from $p$ to $q$ of length $\le (1+\eps)r$.

Let $p,q\in P$ be any pair of points with $\|p-q\|_2 \leq r$. Let's compute the probability of connecting $q$ to $p$ with a path in one round.

Suppose $p$ is partitioned into group $P_i$.
We are going to calculate the probability that $p\in P_{i,j}$ and $q\in V_{i,j}$, for some $j\in [T]$.
Define $F := \{ f\in P_i \mid \|q-f\|_2 > (1+\eps)r \}$ as the set of far points of $q$. 
If $F\cap A_j=\emptyset$,  while $q$ is included in $B_j$, then $q$ will pass the FNS test and thus be included in $V_{i,j}$.

By Lemma~\ref{lem:lsh} and the choice of our parameters ($c_1\leftarrow 1+\eps$ and $n\leftarrow Km$), we have for all $j\in[T]$,
\[
\Pr[p\in A_j \mid q\in B_j] = (Km)^{-o(1)}.
\]
and 
\[
\Pr[f\in A_j \mid q\in B_j] = 1/(Km).
\]

Since $|F| \leq |P_i| = m$, by union bound on all $f\in F$, we can bound the probability that the bad event happens:
\[
\Pr[\exists f\in F,\text{~s.t.~} f\in A_j\mid q\in B_j] \leq 1/K.
\]

Thus, by choose a proper $K = n^{o(1)}$, we have
\[
\Pr[q\in V_{i,j} \textrm{~and~} p\in A_j \mid q\in B_j] \geq (Km)^{-o(1)} - 1/K = (Km)^{-o(1)}(1-o(1)).
\]

Also, By Lemma~\ref{lem:lsh}, we have
\[
|\{j\in [T] : q\in B_j\} | = (Km)^{o(1)} \geq 1.
\]

Thus, with $\geq n^{-o(1)}$ probability, $p\in P_{i,j}$ and $q\in V_{i,j}$ for some $i,j$, and hence the spanner will have a 2-hop path from $q$ to $p$, via $s_{ij}$, with a path of length $(1+3\eps)r$.
By repeating $n^{o(1)}$ rounds, $q$ connects to $p$ with high probability.

\textbf{Bounding the number of edges.}
In one round, the total number of edges added from $S$ to $P$ is at most $\sum_{i,j}|P_{i,j}| \leq n^{5/4+O(\eps)}$, where the last step is by Observation~\ref{obs:alg1}.
So the total number of edges from $S$ to $P$ is at most $n^{5/4+O(\eps)}$.

On the other hand, the total number of edges from $Q=P$ to $S$ is at most
\[
n^{o(1)}\cdot \sum_{i,j}|V_{i,j}| \leq n^{o(1)}\frac{n}{m}\cdot \sum_j |Q_j|=n^{2+o(1)}/m = n^{2-\Omega(\eps^2)},
\]
where the first step is because $V_{i,j} \subseteq Q_j$, the second step is by $\sum_j |Q_j| = n^{1+o(1)}$ (Observation~\ref{obs:alg1}).

Thus, $|E|\leq n^{2-\Omega(\eps^2)}$.

\textbf{Running time.}
By Lemma~\ref{lem:lsh}, we can find all $A_j$ that contains $p$ for all $p\in P$ in $n\cdot (Km)^{1/\eps^2+O(1/\eps)}=n^{5/4+O(\eps)}$ time,
and all $B_j$ that contains $q$ for all $q\in P$ in $n\cdot (Km)^{o(1)} = n^{1+o(1)}$ time.

The major running time cost is on building and querying the FNS data structure on every $P_{i,j}$. We use Lemma~\ref{lem:fns} with $\rho_P = 2/\eps^2$ and $\rho_Q = 0$ (See line~\ref{line:alg1_fns}).

In the preprocessing phase, we spend time $\sum |P_{i,j}|^{\rho_P}$ to build the FNS data structure for every $P_{i,j}$. The total time cost is
\[
\sum_{i,j} |P_{i,j}|^{\rho_P}\leq (\sum |P_{i,j}|)\cdot \max |P_{i,j}|^{\rho_P} \leq n^{5/4+O(\eps)}\cdot m^{\rho_P} = n^{7/4+O(\eps)},
\]
where the second step is because $\sum |P_{i,j}| = n^{5/4+O(\eps)}$, and $|P_{i,j}|\leq |P_i| =m$, the last step is by our choice of $m=n^{\eps^2/4}$.

In the query phase, for every point $q\in P$, if $q$ is in $Q_j$, $q$ needs to query FNS data structure built on $P_{i,j}$ for every $i\in[n/m]$.
Since $q$ is in at most $n^{o(1)}$ different $Q_j$'s, with our choice of $\rho_q=0$, the query time cost on each point $q\in P$ is at most $\sum_{i\in [n/m], j:q\in Q_j}|P_{i,j}|^{\rho_Q} = n^{1+o(1)}/m$. Thus, the query time for all $q\in P$, and also the overall running time is
\[
n\cdot n^{1+o(1)}/m = n^{2-\Omega(\eps^2)}.
\]
\end{proof}

\section{Undirected Non-metric Steiner Spanners}\label{sec:undirected}
In this section, we show how to turn the directed spanner from Theorem~\ref{thm:directed_spanner} into an undirected spanner, with some loss in its size. Here is the formal statement:

\begin{theorem}[Restatement of Theorem~\ref{thm:ubUndir_intro}]\label{thm:ubUndir}
    Fix $\eps\in(0,1)$. For any dataset $P\in \R^d$, we can construct an undirected graph $H$ on vertices $P\cup S$, such that for any $p,q\in P$, 
    $$
    \|p-q\|_2\le \d_H(p,q) < (1+\eps) \cdot \|p-q\|_2.
    $$
    Furthermore, the construction time, the number of edges of $H$ and the size of $S$ are bounded by $O(n^{2-\Omega(\eps^3)}\log \Delta)$, where $\Delta$ is the aspect ratio.
\end{theorem}

The proof of this theorem starts with the directed spanner constructed in the previous section, modifying it further. We note that the directed edges appear in  line~\ref{line:create_star_start} to line~\ref{line:create_star_end} in Algorithm~\ref{alg:directed}, where we link all $q\in V_{i,j}$ to Steiner point $s_{ij}$ with edge length $(1+\eps)r$ and link $s_{ij}$ to all $p\in P_{i,j}$ with edge length $0$. If these directed edges lose their direction, they will potentially create shortcuts between $p,p'\in P_{i,j}$ or $q,q'\in V_{i,j}$ (depending on how we would assign the weights to these edges).

We show how to replace the above (directed) gadget with a more involved (undirected) graph. For now, we focus on one $(V_{i,j},P_{i,j})$ pair. The high level idea is to partition $V_{i,j}$ and $P_{i,j}$ into clusters of bounded diameter $a$ and $b$, respectively, with $a+b = 2(1+\eps)r$. Then, for each pair of clusters $V_{ij}'\subset V_{i,j},P_{ij}'\subset P_{ij}$, we add a Steiner node with connecting to each $q\in V_{i,j}'$ with edge of length $a/2$, and to $P_{i,j}'$ with edge length $b/2$. In this way, we connect $V_{i,j}$ and $P_{i,j}$ with path of length $a/2+b/2=(1+\eps)r$ without shortcuting any pair of points. As long as the number of clusters is $\ll |V_{ij}|,|P_{ij}|$, the number of edges is $\ll |V_{i,j}|\cdot |P_{i,j}|$.

We start with some definitions.

\begin{definition}[Average square distance]\label{def:avg}
For a point set $A$, define $\avg(A) = \sqrt{\E_{p,q\in A}[\|p-q\|_2^2]}$.
Define $\avg_{\max}(A) = \max_{S\subseteq A}\avg(S)$.
\end{definition}
\begin{remark}
When $S$ only has two points $x,y$, $\E_{p,q\in S}[\|p-q\|_2^2] \neq \|x-y\|_2^2$, since there is half probability that $p=q=x$ or $p=q=y$. Thus, $\avg_{\max}^2(A)$ is can be smaller than $\max_{p,q\in A}\|p-q\|_2^2$.
\end{remark}

The average square distance has the following properties.

\begin{claim}
\label{clm:biclique}
    If two sets of points $A,B$ satisfy $\max_{p\in A,q\in B}\|p-q\|_2\le r$, then $\avg(A) + \avg(B) \le 2r$.
\end{claim}
\begin{proof}
    The proof follows from non-embeddability of the $K_{n,n}$ metric into the square of $\ell_2$. In particular, consider the following negative-type inequality applied to $A,B$ (see, e.g., \cite{seidel1991quasiregular} or \cite{schoenberg1935remarks}): for any real $a_p,b_q$ with $\sum_{p\in A} a_p+\sum_{q\in B} b_q=0$, we must have:
    $$
    \sum_{i,j\in A} a_ia_j \|p-q\|_2^2 + \sum_{i,j\in B} b_ib_j \|p-q\|_2^2
    -2\sum_{i\in A,j\in B} a_ib_j \|p-q\|_2^2\le 0.
    $$
    Now set $a_p=1/|A|$ and $b_q=-1/|B|$. The inequality from above becomes $\avg^2(A) + \avg^2(B)\le 2 \E_{p\in A, q\in B} [\|p-q\|_2^2]\le 2r^2$. Thus, $\avg(A) + \avg(B) \leq \sqrt{2\cdot 2r^2}=2r$.
\end{proof}
Note that Claim~\ref{clm:biclique} is also true for any subset of $A$ and $B$. Thus, we have the following proposition.
\begin{proposition}\label{pro:biclique}
If two sets of points $A,B$ satisfy $\max_{p\in A,q\in B}\|p-q\|_2\le r$, then $\avg_{\max}(A) + \avg_{\max}(B) \le 2r$.
\end{proposition}

\begin{definition}[Low diameter decomposition]\label{def:diameter_decomposition}
Let $A$ be a point set. We say $A_1,\cdots, A_T$ is an \emph{$r$-diameter decomposition} of $A$, if they form a partition of $A$, and every $A_i$ has diameter at most $r$.
\end{definition}

\begin{lemma}\label{lem:gadget}
Fix $r,\eps>0$. Suppose we have two point sets $A, B$ with $\max_{p\in A,q\in B}\|p-q\|_2\le r$. Let $x,y > 0$ be any real number such that $x + y \leq 2(1+\eps)r$. Given $A_1,\cdots, A_k$, an $x$-diameter decomposition of $A$, and $B_1,\cdots B_l$, a $y$-diameter decomposition of $B$, we can build an undirected graph $H$ such that for $\d_H$ the shortest path distance in $H$:
    \begin{itemize}
        \item for all $p \in A$, $q\in B$, $\d_H(p,q)\le (1+\eps)r$;
        \item for all $p,q\in A\cup B$, $\d_H(p,q)\ge \|p-q\|_2$,
    \end{itemize}
and the size of $H$ is $O(|A|\cdot l + k\cdot |B|)$. The construction time is $O(|H|)$.
\end{lemma}
\begin{proof}
Choose any $x_0 \geq x$ and $y_0\geq y$ such that $x_0 + y_0 = 2(1+\eps)r$.

For every $A_i$ and $B_j$, we create a Steiner point $s_{i,j}$. For every $p\in A_i$, add edge $(p,s_{i,j})$ with length $x_0/2$. For every $q\in B_j$, add edge $(q,s_{i,j})$ with length $y_0/2$. Now, every $p_1,p_2\in A_i$ are connected with distance $x_0$, every $q_1,q_2\in B_j$ are connected by $y_0$, and every $p\in A_i,q\in B_j$ are connected by $(x_0+y_0)/2 = (1+\eps)r$. Because $A_i$ has diameter at most $x\leq x_0$, $B_j$ has diameter at most $y\leq y_0$, the constraint $\d_H(p,q)\geq \|p-q\|_2$ always holds throughout this procedure.

The number of edges is at most $\sum_{i\in[k],j\in [l]} (|A_i| + |B_j|) = |A|\cdot l + k\cdot |B|$.
\end{proof}

With the above lemma, it remains to show how to construct small low-diameter decompositions, which we will do next.

\subsection{Computing low diameter decomposition}

We now show that we can compute efficient low-diameter decompositions to be used in Lemma~\ref{lem:gadget}. We note that we use $x\approx \avg_{\max}(A)$ and $y\approx \avg_{\max}(B)$. Hence we need an efficient algorithm to compute $\avg_{\max}(A)$-diameter decomposition for any point set $A$. We will use the notions of \emph{$r$-close} and \emph{$r$-far}. Namely, if two points $p,q$ have distance $\leq r$, we say they are a \emph{$r$-close} pair. If their distance is $>r$, we say they are a \emph{$r$-far pair}.

This section gives an important subroutine for the decomposition, presented as Algorithm~\ref{alg:undirected}. 
The input set $P$ has the guarantee that at least $|P|^{2-t}$ pairs of points are $a$-close. The algorithm will then output disjoint clusters $P_1,\cdots,P_T\subseteq P$, in which every $P_i$ has diameter $(1+\eps)a$, and $\cup P_i$ contains a significant portion of $P$. $c$ is a parameter the algorithm can chose from $(0,0.2]$.

\begin{algorithm}[h]\caption{}\label{alg:undirected}
\begin{algorithmic}[1]
\Function{Extract Clusters}{$P\subseteq S^{d-1}$, $a > 0$, $\eps > 0$, $c > 0$, $t>0$}
\State \Comment{Input guarantee: at least $|P|^{2-t}$ pairs of points are $a$-close.}
\State Let $n=|P|$.
\State Let $h$ be a random $(a, (1+\eps)a, p_1, p_2)$-LSH function, with $p_1 = n^{-c}$, $p_2 = n^{-(1+\eps)c}$. \label{line:pick_hash_function}
\State Partition $P$ into $1/p_2$ buckets using $h$. 
\State \Comment{See the proof of Lemma~\ref{lem:cluster_extraction} for the reason why there are $\le 1/p_2$ buckets.}
\State For each bucket $i$, let $C_{h,i}$ be the number of $a$-close pairs in this bucket. Similarly, let $F_{h,i}$ be the number of $(1+\eps)a$-far pairs.
\If{$\exists i$, such that $C_{h,i} \geq \frac{p_1p_2}{4}n^{2-t}$ and $F_{h,i}/C_{h/i} \leq 16\frac{p_2}{p_1}n^t$}\label{line:if}
\State Let $B_i$ be all points in bucket $i$.
\Repeat\label{line:repeat}
    \State randomly pick $k=n^{(\eps c -t)/2}/10$ points from $B_i$.
    \If{they are pairwise $a$-close pairs}
        \State form a cluster with them and delete them from $B_i$ \label{line:extract}
    \EndIf
    \Until{formed $C_{h,i}/(10nk)$ clusters.}
\Else \State Go back to Line~\ref{line:pick_hash_function}
\EndIf
\State \Return all formed clusters
\EndFunction
\end{algorithmic}
\end{algorithm}

The guarantees of the Algorithm~\ref{alg:undirected} are captured by the following lemma.

\begin{lemma}[cluster extraction]\label{lem:cluster_extraction}
Fix an $n$-point set $P\subset \R^d$. Let $\eps \in (0,2)$ be the approximation parameter. Let $a > 0$ be a parameter such that there are at least $n^{2-t}$ $a$-close pairs, where $t=O(\eps)$.

For any parameter $c \in (0, 0.2]$ with $\eps c > t$, Algorithm~\ref{alg:undirected} outputs disjoint clusters each with size $\Omega(n^{(\eps c - t)/2})$. Each cluster has diameter $(1+\eps)a$, i.e., all its points are $(1+\eps)a$-close. Moreover, the total number of points in all clusters is $\Omega(n^{1-2c-\eps c -t})$. Algorithm~\ref{alg:undirected} can be implemented to run in $O(n^{1+c+t})$ expected time.
\end{lemma}
\begin{proof}
In our algorithm, we pick a random $(a,(1+\eps)a,p_1,p_2)$-LSH function with $p_1 = 1/n^c$ and $p_2 = 1/n^{(1+\eps)c}$. We note that an $(a,(1+\eps)a,p_1,p_2)$-LSH function maps an $a$-close pair into the same bucket with probability $\geq p_1$, and map a $(1+\eps)a$-far pair into the same bucket with probability $\leq p_2$.

We can assume, wlog, that the number of buckets is at most $B=1/p_2$, since otherwise, we can randomly group buckets into $1/p_2$ buckets, increasing $p_2$ by at most a factor of $2$. 
The expected number of $(1+\eps)a$-far pairs that collide is $\chi_F\le p_2 n^2$, and the expected number of colliding $a$-close pairs is $\chi_C\ge p_1 n^{2-t}$.

For any such LSH function $h$, we denote $F_h$ and $C_h$ as the number of colliding far and close pairs, respectively. We use $\one{A}=1$ if $A$ is true, otherwise $\one{A}=0$.

Since $\E[C_h]=\chi_C$, we have
\[
\E[C_h\cdot \one{C_h\geq \frac{1}{2}\chi_C}] \geq \frac{1}{2}\chi_C.
\]
Also,
\[
\chi_F=\E[F_h]\ge
\E[C_h\cdot \frac{F_h}{C_h}\cdot \one{\frac{F_h}{C_h}\geq  \frac{4\chi_F}{\chi_C}}]
\ge
\E[C_h\cdot \one{C_h\geq \frac{1}{2}\chi_C \text{~and~} \frac{F_h}{C_h}\geq \frac{4\chi_F}{\chi_C}}]\cdot \frac{4\chi_F}{\chi_C}.
\]
Combining the above two inequalities, we get
\[
\E[C_h\cdot \one{C_h\geq \frac{1}{2}\chi_C \text{~and~} \frac{F_h}{C_h}\leq \frac{4\chi_F}{\chi_C}}]\geq \frac{1}{2}\chi_C - \chi_F\cdot \frac{\chi_C}{4\chi_F} \geq \frac{1}{4}\chi_C.
\]

Since $C_h\leq n^2$, we get
\[
\Pr[C_h\geq \frac{1}{2}\chi_C \text{~and~}C_h/F_h \geq \frac{1}{4}\chi_C/\chi_F] \geq \frac{1}{4}\chi_C/n^2.
\]

Furthermore, there must exist some bucket $i$ such that: 1) the number of close pairs colliding in bucket $i$ is $C_{h,i}\ge C_h/2B$, and 2) the number $F_{h,i}$ of colliding far pairs satisfies that $C_{h,i}/F_{h,i}\ge \frac{1}{4} C_h/F_h$.

This means, with probability $\geq \frac{1}{4}\chi_C/n^2 \geq p_1n^{-t}/4$, we  succeed at the if-check in line~\ref{line:if}, because we have a bucket with $C_{h,i}\ge C_h/2B\ge \tfrac{p_1p_2}{4}n^{2-t}$ and $F_{h,i}/C_{h,i}\le 16\chi_F/\chi_C\le 16 \frac{p_2}{p_1}n^{t}$. Pick $k = \sqrt{p_1n^{-t}/(160p_2)} = \Theta(n^{(\eps c-t)/2}) \geq 1$ points at random from this bucket $i$, and call them $A$. Then, for set $F$ defined as the $(1+\eps)a$-far pairs in $B_i$, we have $\E[|A^2\cap F|]\le 16 \frac{p_2}{p_1}n^t\cdot k^2 < 1/10$. Thus, with $\geq 9/10$ probability, $k$ points are pairwise $(1+\eps)a$-close and they are extracted in line~\ref{line:extract}. Note that each extraction will decrease $C_{h,i}$ by at most $nk$, so we can extract $C_{h,i}/(10nk)$ times safely as the same analysis holds.

\textbf{Running time.} We  check line~\ref{line:if} for $O(n^t/p_1)$ times in expectation. Each time we use $O(n)$ time to implement the hashing. In each check, instead of using $n^2$ time calculating exact $C_{h,i}$ and $F_{h,i}$, we can do sampling to estimate. $O(\frac{n^t\log n}{p_1p_2})$ samples can $1+o(1)$ approximate $C_{h,i}$ w.h.p, and $O(\frac{\log n}{p_2^2})$ samples is enough for $F_{h,i}$.

Total running time would be 
\begin{align*}
O\left(\frac{n^t}{p_1}\Big(n+B\big(\frac{n^t\log n}{p_1p_2} + \frac{\log n}{p_2^2}\big) \Big) + \frac{C_{h,i}k^2}{10nk} \right)
= & ~ O(n^{1+c+t} + n^{4c+2\eps c + 2t + o(1)} + n^{4c+3\eps c+t+o(1)} +n^{1-2c})\\
= & ~ O(n^{1+c+t}).
\end{align*}
\end{proof}

\subsection{Proof of Theorem~\ref{thm:ubUndir}}
The last step is to compute 
small-diameter decomposition for sets $V_{ij},P_{ij}$. In particular, we will use Algorithm~\ref{alg:undirected} to compute
$(\avg_{\max}(A) + \eps\cdot \dia(A))$-diameter decomposition of a set $A$, where $\dia(A)$ is the diameter of $A$. 

We accomplish this in the next three lemmas. Lemma~\ref{lem:avg_dis_square} shows that most pairs inside a set $A$ are at distance at most $\avg(A)$.
Lemma~\ref{lem:compute_dec} shows a decomposition of a set $A$ into clusters with diameter $\le(\avg_{\max}(A) + \eps\cdot \dia(A))$.  Lemma~\ref{lem:compute_dec_v} shows how to decompose  specifically  
 $V_{i,j}$ from Algorithm~\ref{alg:directed}.

\begin{lemma}\label{lem:avg_dis_square}
Given a point set $A$ and $\eps \in (0,1/2)$, let the average distance be define in Def.~\ref{def:avg}, i.e., 
$\avg(A)=\sqrt{\E_{p,q\in A}[\|p-q\|_2^2]}$. Then, there are at least $|A|^2\eps$ pairs of $p,q\in A$ such that $\|p-q\|_2\leq (1+\eps)\avg(A)$.
\end{lemma}
\begin{proof}
By Markov's, $\Pr_{p,q\in A}[\|p-q\|_2^2 > (1+\eps)^2\avg^2(A)] \leq \frac{\avg^2(A)}{(1+\eps)^2\avg^2(A)}\leq 1-\eps$. Thus, there are $\eps$ fraction pairs of $p,q\in A$ that have $\|p-q\|_2\leq (1+\eps)\avg(A)$.
\end{proof}

\begin{lemma}\label{lem:compute_dec}
Fix $\eps >0$. Consider a set $A$ of size $n > \omega(1/\eps)$. We can construct $A_1,\cdots,A_T$, which is a $(\avg_{\max}(A) + \eps\cdot \dia(A))$-diameter decomposition of $A$ in  $O(n^{1.4})$ time. Furthermore, we have $T\leq n^{1-\Omega(\eps)}$.
\end{lemma}
\begin{proof}
Since every $p,q\in A$ has $\|p-q\|_2\leq \dia(A)$, we can compute a quantity $\hat{\avg}(A)$ in $\poly(1/\eps, \log n)$ time satisfying
$\avg(A) \leq \hat{\avg}(A) \leq \avg(A) + \eps \cdot \dia(A)$ (simply by computing an empirical average of a number of sampled pairs).

By Lemma~\ref{lem:avg_dis_square}, there are at least $n^2\eps$ number of $(1+\eps)\hat{\avg}(A)$-close pairs. Let $t=o(1)$ so that $n^2\eps =n^{2-t}$. Fix $c=0.05$. Run $\textsc{Extract Clusters}(A, a\leftarrow (1+\eps)\hat{\avg}(A), \eps, c\leftarrow 0.05, t)$. By Lemma~\ref{lem:cluster_extraction}, we extract $\Omega(n^{1-2c-\eps c -t})$ points from $A$ and that takes $O(n^{1+c+t})$ time.

Denote the remaining part of $A$ as $A'$.
Now, iteratively recalculate $\hat{\avg}(A')$ and run $\textsc{Extract Clusters}$ on $A'$ until no points are remained.
Denote $A_1,\cdots,A_T$ as the final partition of $A$. By Lemma~\ref{lem:cluster_extraction}, $T$ is at most $\log n\cdot \frac{n}{n^{(\eps c-t)/2}} = n^{1-\Omega(\eps)}$ since each $A_i$ has size lower bound. And every $A_i$ has diameter at most $(1+\eps)(\avg_{\max}(A) + \eps\cdot \dia(A))\leq \avg_{\max} + 3\dia(A)$. The total running time is $O(\log n\cdot \frac{n}{n^{1-2c-\eps c-t}}n^{1+c+t}) = O(n^{1.4})$. Replace $\eps$ with $\eps/3$ to conclude the lemma.
\end{proof}

\begin{lemma}[Decompose $V_{i,j}$]\label{lem:compute_dec_v}
Suppose we are in $\textsc{Build Spanner}(P,r,\eps)$ (Algorithm~\ref{alg:directed}) and let $n$, $Q_j$, $V_{i,j}$ be defined in Algorithm~\ref{alg:directed}. Using $n^{2-\Omega(\eps^2)}$ time, we can construct $(\avg_{\max}(V_{i,j}) + 3\eps r)$-diameter decomposition of $V_{i,j}$ with at most $|Q_j|^{1-\Omega(\eps)}$ clusters for every $V_{i,j}$.
\end{lemma}
\begin{proof}
The idea is to compute low-diameter decomposition for all of $Q_{j}$, and the decomposition for $V_{i,j}$ can be calculated in a straightforward way. 

Fix $Q_j$. For every $r_0 \in \{0,\eps r,2\eps r,\cdots, 2r\}$, we iteratively run Algorithm~\ref{alg:undirected} with parameters \textsc{Extract Clusters}($Q_j$, $a\leftarrow r_0$, $\eps\leftarrow \eps$, $c\leftarrow 0.1$, $t\leftarrow 0.1\eps$) and delete the extracted points from $Q_j$, until we cannot extract anymore --- specifically, until the Repeat loop (line~\ref{line:repeat}) cannot be satisfied within the bounded time Lemma~\ref{lem:cluster_extraction} gives, which can be checked directly. Let the extracted clusters be $A_1^{r_0},\cdots A_T^{r_0}$ ($T\leq |Q_j|^{1-\Omega(\eps)}$) and the remaining set be $A_{\textrm{re}}^{r_0}$ ($|A_{\textrm{re}}^{r_0}|\leq |Q_j|$). Note that every $A^{r_0}_i$ has diameter $r_0$.

Now consider one $V_{i,j}$. Let $\avg_{\max}(V_{i,j}) = \max_{S\subseteq V_{i,j}}\avg(S)$ be defined in Def.~\ref{def:avg}. Because every point in $V_{i,j}$ is $r$-close to $P_{i,j}$, we have $\avg_{\max}(V_{i,j})\leq 2r$. We claim that there exists $r_0 \leq \avg_{\max}(V_{i,j}) + 2\eps r$ such that the  partition of $V_{i,j}$ defined as $\{A_1^{r_0}\cap V_{i,j},\cdots, A_T^{r_0}\cap V_{i,j}\} \cup \{v\}_{v\in A^{r_0}_{\textrm{re}} \cap V_{i,j}}$ consists of at most $|Q_j|^{1-\Omega(\eps)}$ clusters, and therefore, this is the decomposition of $V_{i,j}$ we want.
The only thing we need to prove is that $|A^{r_0}_{\textrm{re}}\cap V_{i,j}| = |Q_j|^{1-\Omega(\eps)}$.

Let $B = A^{r_0}_{\textrm{re}}\cap V_{i,j}$.  
By Lemma~\ref{lem:avg_dis_square}, there are at least $|B|^2\eps$ pairs of $p,q\in B$ with $\|p-q\|_2\leq (1+\eps)\avg(B)$. Let $r_0\geq (1+\eps)\avg(B)$ be the least multiplicity of $\eps r$. So $r_0\leq \avg_{\max}(V_{i,j})+3\eps r$. Since $A^{r_0}_{\mathrm{re}}$ as a remaining set fails on $\textsc{Extract Clusters}(A^{r_0}_{\mathrm{re}},r_0,\eps,c \leftarrow 0.1, t\leftarrow 0.1\eps)$, by Lemma~\ref{lem:cluster_extraction}, this means the number of $r_0$-close pairs is at most $|A^{r_0}_{\mathrm{re}}|^{2-t}\leq |Q_j|^{2-0.1\eps}$. This means $|B|^2\eps \leq |Q_j|^{2-0.1\eps}$. Therefore, $|B| = |Q_j|^{1-\Omega(\eps)}$.

There are at most $O(1/\eps)$ different $r_0$ and we can try all to find the correct one. Use Lemma~\ref{lem:compute_dec}, computing the decomposition for all $Q_j$ and $r_0$ uses at most $\frac{1}{\eps}\sum_j |Q_j|^{1.4} \leq n^{1.5}$ running time.
Thus, decomposing all $V_{i,j}$ is done in time
\[
O(n^{1.5}+ \frac{1}{\eps}\sum_{i,j}|V_{i,j}|) \leq n^{2-\Omega(\eps^2)}.
\]
\end{proof}

\begin{proof}[Proof of Theorem~\ref{thm:ubUndir}]
Suppose we are in $\textsc{Build Spanner}(P,r,\eps)$ (Algorithm~\ref{alg:directed}). Let $Q_j$, $V_{i,j}$, $P_{i,j}$ be as in the algorithm. Now, instead of doing line~\ref{line:create_star_start} to line~\ref{line:create_star_end}, we do the following. 

First, compute an $(\avg_{\max}(P_{i,j}) + 2\eps\cdot r)$-diameter decomposition of $P_{i,j}$, denoted as $A_1,\cdots,A_k$, and compute an $(\avg_{\max}(V_{i,j}) + 3\eps\cdot r)$-diameter decomposition of $V_{i,j}$, denoted as $B_1,\cdots,B_l$. 
Then, use the gadget in Lemma~\ref{lem:gadget} to add edges in our undirected spanner.
The reason we can use the gadget is because $\avg_{\max}(V_{i,j}) + \avg_{\max}(P_{i,j})\leq 2r$ by Proposition~\ref{pro:biclique}. Thus pairs in $V_{ij}\times P_{ij}$ are distorted by at most $1+O(\eps)$.

Decomposing all $P_{i,j}$ can be done in running time proportional to
\[
\sum_{i,j} |P_{i,j}|^{1.4} \leq (\sum_{i,j} |P_{i,j}|)^{1.4} \leq n^{1.25\cdot 1.4} = n^{1.75}.
\] 
Computing the decompositions of all $V_{i,j}$ can be done in $n^{2-\Omega(\eps^2)}$ time.

\textbf{Size of spanner.} By Lemma~\ref{lem:gadget}, every pair of $V_{i,j}$ and $P_{i,j}$ adds $|V_{i,j}|\cdot k + l\cdot |P_{i,j}|$ edges to the spanner.
We have $k\leq |P_{i,j}|^{1-\Omega(\eps)}$ (Lemma~\ref{lem:compute_dec}) and $l\leq |Q_j|^{1-\Omega(\eps)}$ (Lemma~\ref{lem:compute_dec_v}). 
Since $V_{i,j}\subseteq Q_j$, the total size of our spanner is at most
\[
\log \Delta \cdot \left( \sum_{i,j} |Q_{j}|\cdot |P_{i,j}|^{1-\Omega(\eps)} + |Q_{j}|^{1-\Omega(\eps)}\cdot |P_{i,j}| \right).
\]

Here we copied the constraints for $Q_{j}$ and $P_{i,j}$ from Observation~\ref{obs:alg1}.
\begin{enumerate}
\item for all $i,j$, $P_{i,j} \subseteq P_i$ and $|P_i|=m$. \label{item:1}
\item $\sum_j |Q_j| = n^{1+o(1)}$; \label{item:2}
\item $\sum_{i,j} \E[|P_{i,j}|] = n^{5/4+O(\eps)}$; \label{item:3}
\item $\forall j, \sum_{i}|P_{i,j}|\leq |P| = n$, \label{item:4}
\item $i\in [n/m]$ and $m = n^{\eps^2/4}$. \label{item:5}
\end{enumerate}

We bound the two terms as follows.

\begin{enumerate}
\item
\begin{align*}
& ~ \sum_{i,j} |Q_{j}|\cdot |P_{i,j}|^{1-\Omega(\eps)}\\
\leq & ~ \sum_{i,j} |Q_j|\cdot (\max_{i,j}|P_{i,j}|^{1-\Omega(\eps)})\\
\leq & ~ \frac{n}{m}\cdot \sum_{j} |Q_j|\cdot m^{1-\Omega(\eps)}\\
\leq & ~ n^2/m^{\eps} = n^{2-\Omega(\eps^3)},
\end{align*}
where the second step is because $i\in [n/m]$ \eqref{item:5} and \eqref{item:1}, the third step is by \eqref{item:2}.
\item 
\begin{align*}
& ~ \sum_{i,j} |Q_j|^{1-\Omega(\eps)}\cdot |P_{i,j}|\\
\leq & ~ \sum_{i,j,|Q_j|\leq \sqrt{n}} |Q_j|^{1-\Omega(\eps)}\cdot |P_{i,j}| + \sum_{i,j,|Q_j|\geq \sqrt{n}} |Q_j|^{1-\Omega(\eps)}\cdot |P_{i,j}|\\
\leq & ~ \sqrt{n}^{1-\Omega(\eps)}\cdot \sum_{i,j}|P_{i,j}| + \sum_{j,~|Q_j|\geq \sqrt{n}} |Q_j|^{1-\Omega(\eps)}\cdot n\\
\leq & ~ n^{1.75 + O(\eps)} + n^{2-\Omega(\eps)},
\end{align*}
where the second step is by $\sum_j |P_{i,j}|=n$ \eqref{item:4}, the last step is by \eqref{item:3}.
\end{enumerate}

Concluding, we proved that the size of our spanner is $n^{2-\Omega(\eps^3)}\log \Delta$. 
\end{proof}

\section{Removing dependence on $\Delta$}\label{sec:remove_delta}

If we ignore the time cost for constructing, we can remove the dependence on $\Delta$ in the number of edges used in the spanners. Below we show this for the directed spanner reducing the size from $n^{2-\Omega(\eps^2)}\log \Delta$ to 
$n^{2-\Omega(\eps^2)}$; similar result for the undirected spanner is analogous. We use the idea from \cite{his13}.

We start by defining the $\delta$-net.
\begin{definition}[$\delta$-net]
Let $(X,\rho)$ be a metric space, and let $\delta > 0$. A maximal set $Y\subseteq X$, such that for any $x,y\in Y$, $\rho(x,y) > \delta$, is called a $\delta-$net for $(X,\rho)$.
\end{definition}

Assume the dataset has closest distance $a$ and largest distance $a\cdot \Delta$. Let $b = a\cdot \eps$.
Define $K=\log_{1+\eps} (\Delta/\eps)$ so that $b\cdot (1+\eps)^K = a\cdot \Delta$.
For every $i\in \{0,1,\cdots, K\}$, let $N_i \subseteq P$ be a $b(1+\eps)^i$-net for $(P,\ell_2)$. This means every point $p$ not in $N_i$ must have a near neighbor in $N_i$ that is $b(1+\eps)^i$ close, and we call this point the leader of $p$, denoted as $L_i(p)$.

We can pick those nets so that
\[
P = N_0 \supseteq N_1 \supseteq N_2 \supseteq \cdots \supseteq N_K.
\]

We suppose to run Lemma~\ref{lem:core} on each net $N_i$ with  parameter $r_i=a(1+\eps)^i$. But note that many point in $N_i$ may not even has a neighbor within distance $r_i$. Thus, we will do the following to reduce the size of the spanner. For $i\in [k]$, 
let $V_{i}$ be the set of points in $N_i$ that has at least one neighbor within distance $r_i$ in $N_i$. Then we run Lemma~\ref{lem:core} on each $V_i$ with parameter $r=r_i$ to get $H_i$. We claim that $\cup_i H_i$ is a spanner with $(1+7\eps)$ dilation.

We prove by induction on the distance $r$ that for every pair of point with $\|p-q\|_2\leq r$, $\d_H(p,q)\leq (1+7\eps)r$. Assume this is true for all pairs of point with distance $<r$.
Let $p,q$ be $\|p-q\|_2 = r\in [a(1+\eps)^{i-4},a(1+\eps)^{i-3}]$. If $p \in N_i$, we let $x = p$, otherwise we let $x = L_i(p)\in N_i$ be the leader of $p$. Similarly, we define $y = q$ if $q\in N_i$ or otherwise we let $y = L_i(q)$. Then $\vec{d}_H(p,q) \leq \vec{d}_H(p,x) + \vec{d}_H(x,y) + \vec{d}_H(y,q)$. Since $\|p-x\|_2,\|q-y\|_2\leq b(1+\eps)^i < r$, by induction we must have $\vec{d}_{H}(p,x),\vec{d}_H(q,y) \leq (1+7\eps)\cdot b(1+\eps)^i \leq \eps(1+11\eps)r$. Since $\|x-y\|_2\leq \|x-p\|_2+\|p-q\|_2+\|q-y\|_2 \leq (1+3\eps) r\leq r_i$ and $x,y\in N_i$, $x,y$ must be also in $V_i$. Thus, we have
\[
\vec{d}_{H_i}(x,y) \leq (1+\eps) r_i \leq (1+4\eps)r.
\]
Thus, $\vec{d}_{H}(p,q)\leq \vec{d}_{H}(p,x) + \vec{d}_{H_i}(x,y) + \vec{d}_{H}(q,y)\leq (1+7\eps) \|p-q\|_2$.

Also, because $\forall p,q\in V_i, \vec{d}_{H_i}(p,q)\geq \|p-q\|_2$, so $\forall p,q\in P, \vec{d}_{H}(p,q)\geq \|p-q\|_2$.

Next, we will upper bound the size of $V_i$.

For every point $p\in P$ and $i\in [K]$, we create an imaginary point called $(i,p)$ if $p\in N_i$. Denote $z(p)$ as the largest integer that $p\in N_{z(p)}$.
We build a tree on all such points as follows. We link an edge between $(i,p)$ to $(i+1,p)$ for all $i\leq z(p)-1$, and link $(z(p),p)$ to $(z(p)+1,q)$ where $q=L_{z(p)}(p)$ is the leader of $p$. The root of the tree is $(K,p^*)$, in which $N_K = \{p^*\}$ has only one point. 

Consider any $i\in [K]$ and $p\in P$ such that $p\in V_i$. This means $p$ has a point $q$ that is $a(1+\eps)^i$ close to $p$. By properties of $\delta$-net, $p$ and $q$ cannot co-exists in net $N_{i+k}$, where $k:=\log_{1+\eps}(b/a)$. This means there must be a vertex on the chain $\{(i,p),(i+1,p),\cdots,(i+k,p)\}$ having two children. 

The tree has $n$ leaves and it has at most $n-1$ vertices having two children, so $\sum |V_i| = O(kn)$.

The size of the spanner $H$ is 
\[
\sum_i |V_i|^{2-\Omega(\eps^2)} = (n/\eps^2)^{2-\Omega(\eps^2)} = n^{2-\Omega(\eps^2)}.
\]

\begin{theorem}[Existence result for directed spanner]
Fix $\eps\in(0,1)$. For any dataset $P\in \R^d$, there exists a directed graph $H$ on vertices $P\cup S$ with size $n^{2-\Omega(\eps^2)}$, such that for any $p,q\in P$, 
    $$
    \|p-q\|_2\le \vec{\d}_H(p,q) < (1+\eps) \cdot \|p-q\|_2.
    $$
\end{theorem}

Using the same technique, we can prove similar results for undirected graph.

\begin{theorem}[Existence result for undirected spanner]
Fix $\eps\in(0,1)$. For any dataset $P\in \R^d$, there exists an undirected graph $H$ on vertices $P\cup S$ with size $n^{2-\Omega(\eps^3)}$, such that for any $p,q\in P$, 
    $$
    \|p-q\|_2\le \d_H(p,q) < (1+\eps) \cdot \|p-q\|_2.
    $$
\end{theorem}

\appendix

\section*{Appendix}

\section{Data-independent LSH from \cite{alrw17}}\label{sec:lsh}

We show here how Lemma~\ref{lem:lsh} follows from results in~\cite[Section 3]{alrw17}. We use the notations from \cite{alrw17}, noting that our $\rho_Q/\rho_P$ corresponds to $\rho_q/\rho_u$ from \cite{alrw17}.

\begin{definition}[Definition of $\alpha(s)$, $\beta(s)$]
For any $0<s<2$, let $u,v$ be any two points on a unit sphere $S^{d-1}$ at distance $s$. 
Define $\alpha(s) := 1-s^2/2$ as the cosine of the angle between $u$ and $v$, and $\beta(s) :=\sqrt{1-\alpha^2(s)}$ as the sine of the same angle.
\end{definition}

\begin{definition}[Definition of $F(\eta)$ and $G(s,\eta,\sigma)$]
Let $u$ be any point on $S^{d-1}$. Define
\[
F(\eta) := \Pr_{z\sim N(0,1)^d}[\langle u, z\rangle \geq \eta].
\]
For any $0<s<2$, let $v$ be any point on $S^{d-1}$ with $\dis(u,v)=s$. Define
\[
G(s,\eta,\sigma) := \Pr_{z\sim N(0,1)^d}[\langle u, z\rangle \geq \eta \wedge \langle v,z \rangle \geq \sigma].
\]

Note that $F(\eta),G(s,\eta,\sigma)$ doesn't depend on the specific choice of $u$ and $v$ due to the symmetry of Gaussian.
\end{definition}

\begin{lemma}[Lemma~3.1 of \cite{alrw17}]\label{lem:f}
For $\eta \rightarrow \infty$,
\[
F(\eta) =e^{-(1+o(1))\cdot \frac{\eta^2}{2}}.
\]
\end{lemma}

\begin{lemma}[Lemma~3.2 of \cite{alrw17}]\label{lem:g}
For every $s \in (0,2)$, if $\eta,\sigma\rightarrow \infty$ and $\alpha(s)\cdot \eta<\sigma<\frac{1}{\alpha(s)}\cdot \eta$, one has
\[
G(s,\eta,\sigma)= e^{-(1+o(1))\cdot \frac{\eta^2 + \sigma^2 - 2\alpha(s)\eta \sigma}{2\beta^2(s)}}.
\]
\end{lemma}

\begin{lemma}[Data-independent LSH from Section~3 of \cite{alrw17}]\label{lem:alrw_1}
Let $r \in (0,2)$ be a fixed parameter. For any integer $K\geq 1$ and $\eta_q,\eta_u >0$, 
there is a data structure that, takes a random seed $s$, implicitly gives two random collections of subspaces $A_1,A_2,\cdots,A_T\subset \R^d$ and $B_1,B_2,\cdots, B_T \subset \R^d$. For any point $p$, denote $\mathcal{A}_p = \{A_i : p \in A_i\}$ as the set of subspaces $A_i$ that contains $p$. Similarly, define $\mathcal{B}_p= \{B_i : p\in B_i\}$.
\begin{enumerate}
\item $\forall p\in S^{d-1}$, $\E_s[|\mathcal{A}_p|] = \Theta(F(\eta_u) / G(r,\eta_u,\eta_q) )^K$ ;
\item $\forall p\in S^{d-1}$, $\E_s[|\mathcal{B}_p|] = \Theta(F(\eta_q) / G(r,\eta_u,\eta_q) )^K$ ;
\item $\forall p,q\in S^{d-1}, i\in [T]$,
\[
\Pr[p\in A_i \mid q\in B_i] =(G(\|p-q\|_2, \eta_u,\eta_q)/F(\eta_q))^K
\]
\item $\forall p\in S^{d-1}$, one can find all $A_i \in \mathcal{A}_p$ in $O(|\mathcal{A}_p|\cdot \log n\cdot 1/G(r,\eta_u,\eta_q))$ time, and all $B_i \in \mathcal{B}_p$ in $O(|\mathcal{B}_p|\cdot \log n\cdot 1/G(r,\eta_u,\eta_q))$ time;
\end{enumerate}
\end{lemma}

The following lemma shows that with properly chosen $\eta_u$, $\eta_q$ and $K$, Lemma~\ref{lem:lsh} can be proven from Lemma~\ref{lem:alrw_1}.

\begin{lemma}[parameters]
Let $r=(\log n)^{-1/8}$ and $c = 1+\eps$ is the approximation ratio.
Then, for any $n = \omega(1)$, there is a choice of $\eta_u,\eta_q>0$ and $1\leq K \leq O(\sqrt{\ln n})$ such that 
\begin{enumerate}
\item $(F(\eta_u)/G(r,\eta_u,\eta_q))^K = n^{1/\eps^2+O(1/\eps)}$;
\item $(F(\eta_q)/G(r,\eta_u,\eta_q))^K = n^{o(1)}$;
\item $(G(cr,\eta_u,\eta_q)/F(\eta_q))^K = 1/n$.
\item 
$1/G(r,\eta_u,\eta_q)=n^{o(1)}$.
\end{enumerate}
\end{lemma}
\begin{proof}
Let $\sigma,\tau$ be two parameters that $F(\eta_u)^K = n^{-\sigma}$ and $F(\eta_q)^K = n^{-\tau}$. Thus, $G(r,\eta_u,\eta_q)$ can be written as $n^{-\frac{\sigma + \tau - 2\alpha(r)\sqrt{\tau \sigma}}{\beta^2(r)}}$.
We set $\sqrt{\tau} := \frac{\beta(c r)}{\alpha(r)-\alpha(c r)}$ and $\sqrt{\sigma} := \alpha(r) \cdot \sqrt{\tau}$ as what \cite{alrw17} did in Section 3.3.3, in the case where $\rho_q = 0$. 

The proof of part 3 follows the same proof as in \cite{alrw17}. Here we have

\begin{align*}
\ln \left(\frac{G(cr,\eta_u,\eta_q)}{F(\eta_q)}\right)^K / \ln (n)
= \left( \tau - \frac{\tau + \sigma - 2\alpha(cr)\sqrt{\tau \sigma}}{\beta^2(cr)}\right)= -1.
\end{align*}

Also, part 2 is true because

\begin{align*}
(F(\eta_q)/G(r,\eta_u,\eta_q))^K = n^{\rho_q + o(1)},
\end{align*}
where 
\begin{align*}
\rho_q = \frac{\sigma + \tau -2\alpha(r)\sqrt{\tau \sigma}}{\beta^2(r)}- \tau =\frac{(\sqrt{\sigma} - \alpha(r) \sqrt{\tau})^2}{\beta^2(r)} =0.
\end{align*}

For part 1, we have
\begin{align*}
(F(\eta_u)/G(r,\eta_u,\eta_q))^K = n^{\rho_u + o(1)},
\end{align*}
where
\begin{align*}
\rho_u = \frac{\sigma + \tau -2\alpha(r)\sqrt{\tau \sigma}}{\beta^2(r)}- \sigma = \frac{(\alpha(r)\sqrt{\sigma}-\sqrt{\tau})^2}{\beta^2(r)} = 1/\eps^2+O(1/\eps) + O(r).
\end{align*}

Part 4 follows by computing $G$ directly and using the fact that $r=(\log n)^{-1/8}$.
\end{proof}

\bibliographystyle{alpha}
\bibliography{ref}
\end{document}